\documentclass[a4paper,11pt]{article}
\usepackage{jheppub} 
\usepackage{lineno}
\usepackage{jheppub} 
\usepackage{lineno}
\usepackage{tikz}
\usepackage{wrapfig}
\usetikzlibrary{arrows.meta}
\usepackage[utf8]{inputenc}
\usepackage[backend=biber,style=numeric-comp,sorting=none]{biblatex}
\usepackage{amsmath}

\usepackage{amsthm}
\usepackage{amssymb}
\usepackage{graphicx}
\usepackage{slashed}
\usepackage{MnSymbol}
\usepackage{mathtools}
\usepackage{cancel}
\usepackage{xcolor}
\usepackage{simpler-wick}
\usepackage{hyperref} 
\hypersetup{backref=true,       
    pagebackref=true,               
    hyperindex=true,                
    colorlinks=true,                
    breaklinks=true,                
    urlcolor= black,                
    linkcolor= blue,                
    bookmarks=true,                 
    bookmarksopen=false,
    filecolor=black,
    citecolor=blue,
    linkbordercolor=blue
}
\numberwithin{equation}{section}
\newtheorem{theorem}{Theorem}
\theoremstyle{definition}

\newcommand{\wh}{\widehat}
\newcommand{\wt}{\widetilde}

\newcommand{\efill}{\;\;\;\;\;\;\;\;\;\;}
\newcommand{\be}{\begin{equation}}

\newcommand{\ee}{\end{equation}}
\newcommand{\bthe}{\begin{theorem}}
\newcommand{\ethe}{\end{theorem}}
\newcommand{\bea}{\begin{eqnarray}}
\newcommand{\eea}{\end{eqnarray}}
\newcommand{\al}{\alpha}
\renewcommand{\d}{\delta}

\newcommand{\la}{\lambda}

\newcommand{\Om}{\Omega}

\renewcommand{\t}{\theta}

\newcommand{\hlf}{\frac{1}{2}}

\newcommand{\M}{\mathcal{M}}

\newcommand{\p}{\partial}

\newcommand{\R}{\mathbb{R}}
\newcommand{\rr}{\rightarrow}

\newcommand{\Z}{\mathbb{Z}}

\newcommand{\lp}{\left(}
\newcommand{\rp}{\right)}
\newcommand{\ls}{\left[}
\newcommand{\rs}{\right]}

\newcommand{\bz}{\bar{z}}

\title{\boldmath Boundaries of the compact free boson through edge modes
and boundary SymTFT}
\author{Daniel Robbins} \author{, Subham Roy}\author{, and Hassaan Saleem
}
\affiliation{Department of Physics, University at Albany,\\ 1400 Washington Avenue, Albany, NY, U.S.A}

\emailAdd{dgrobbins@albany.edu, sroy3@albany.edu, hsaleem@albany.edu}

\abstract{We study boundary conditions of the compact free boson theory by coupling the bulk scalar to one-dimensional topological edge modes. We show that a single compact edge mode realizes the Dirichlet family, while a two field system realizes the Neumann family; more general compact edge theories can describe superpositions of boundary conditions and reproduce the expected boundary operator spectra. We then study this system in the framework of boundary SymTFT. We show that the edge mode theories of Dirichlet and Neumann boundaries are encoded in the degrees of freedom localized on the corners and the choice of polarization on the topological face of the boundary SymTFT. We further show that the T-duality defect exchanges the data associated with the Dirichlet and Neumann configurations. Finally, we explore extensions involving non-compact edge modes and multiple bulk fields, obtaining continuous boundary spectra in the first case and mixed boundary conditions in the second.}
\begin{document}
\maketitle
\flushbottom
\section{Introduction}

To define quantum field theories (QFT) on a manifold with boundary, we must specify boundary conditions to ensure that the boundary variation cancels and ensure that we have a well-posed variational problem. Ordinarily, this is done by restricting the boundary values of the fields and their derivatives. Alternatively, certain boundary conditions can be implemented by coupling the bulk theory to additional boundary degrees of freedom, e.g., adding a boundary TFT \cite{Arbalestrier:2025jsg, Kapustin:2010if, Kapustin:2009av, Cordova:2023ent, Banerjee:2026ywy, Banerjee:2026wop}. Recently, this technique was used to impose boundary conditions on $4d$ Maxwell theory (with the $\theta$-term) defined on a $ 4$-manifold $\mathcal{M}_4$ with boundary \cite{Arbalestrier:2025jsg}.

In this paper, we use the same techniques to revisit one of the simplest and best-understood boundary conformal field theories: the compact free boson. We focus on this example because its Dirichlet and Neumann boundaries, their continuous families, and their exchange under T-duality are well known. Thus, the compact free boson provides a controlled setting in which we can apply this edge mode technique. We first identify the one-dimensional topological theories which reproduce the canonical boundary conditions. We observed that the introduction of a single compact edge field produces the relevant characteristics of Dirichlet boundary condition, whereas a system of two compact edge fields produces a Neumann boundary. Adding a one-dimensional $\theta$-angle type terms to the respective boundary action then generates the full families of Dirichlet and Neumann boundary conditions \cite{Oshikawa:1996, Choi:2023xjw}.

As the next step, we generalize the edge mode action by introducing $N$ edge degrees of freedom. We show that having an odd (even) number of edge modes coupled non-degenerately eventually imposes Dirichlet (Neumann) boundary conditions. This general edge mode action can also describe a superposition of multiple $D0$ branes on the target space (or dual target space) circle. We also confirm, by calculating the OPEs of the holomorphic stress tensor with the vertex operators built out of the edge modes, that the spectrum of the conformal weights of these vertex operators matches the expected spectrum of boundary fields for the Dirichlet and Neumann boundaries.

This bulk--boundary construction, however, opens the question, how should T-duality act on these additional degrees of freedom? To answer this question we resort to the  framework of boundary symmetry topological field theory (SymTFT) \cite{Bhardwaj:2024igy, Copetti:2024onh}, which is a generalization of the SymTFT framework to capture the symmetries of theories defined on non-closed manifolds. 

The modern point of view, in which symmetries are identified with topological defects, goes back to the seminal paper \cite{Gaiotto:2014kfa} (for review, see \cite{Costa:2024wks,Iqbal:2024pee, McGreevy:2022oyu, Brennan:2023mmt, Shao:2023gho, Bhardwaj:2023kri, Schafer-Nameki:2023jdn, Bah:2022xfv, Cordova:2022ruw, Freed:2022iao}, and the references therein.). One of the tools that came out of this program is the Symmetry topological field theory (SymTFT) \cite{Apruzzi:2021nmk, Freed:2012bs, Kaidi:2022cpf, Kaidi:2023maf, Lin:2025oml, Freed:2022qnc, Gaiotto:2020iye, Bhardwaj:2023bbf, Baume:2023kkf, Bhardwaj:2023ayw, Cvetic:2024dzu, DelZotto:2024tae}. Abstractly, SymTFT is a $(d+1)$-dimensional topological field theory, which when placed on a slab,  can separate the symmetry data of a $d$-dimensional QFT from its dynamics. One end of the slab carries a topological boundary condition, which encodes the symmetry. It decides which topological operators of the bulk can end on it and these eventually become the operators charged under the symmetry, and the objects dual to these are identified with symmetry generators. The other end of the slab carries a non-topological boundary condition, which encodes the dynamics of the QFT. Shrinking the slab, i.e. colliding the two boundaries, recovers the the partition function of the $d$-dimensional QFT. Originally formulated for finite symmetries, this construction has been extended to continuous symmetries \cite{Antinucci:2024zjp, Brennan:2024fgj, Bonetti:2024cjk, Bonetti:2025dvm, Hasan:2024aow, Gagliano:2024off, Arbalestrier:2025poq, Apruzzi:2025hvs, Jia:2025vrj, Bah:2026gcf}.

When the $d$-dimensional theory itself lives on a manifold with boundary, the SymTFT slab (also known as the sandwich) acquires a third boundary component, a topological face which stretches between the two boundaries of the slab, producing two corners where this topological face meets the symmetry boundary and the physical boundary (see Figure~\ref{fig:compact-boson-boundary-symtft} for an illustration.). This construction is known as the boundary SymTFT \cite{Bhardwaj:2024igy, Copetti:2024onh, Arbalestrier:2025jsg, Bonetti:2025dvm, GarciaEtxebarria:2024jfv}. The face carries a topological boundary condition for the bulk theory, and the corners (in general) can carry additional degrees of freedom, whose role is to make the condition on the face compatible with the one on the boundaries of the SymTFT. After the slab is collapsed, the face and the corners together produce the boundary condition of the $d$-dimensional theory. 

This is exactly the structure we use in this paper. We will see that when applied to the compact free boson theory, the boundary condition of the free boson theory is encoded in a choice of the the symmetry boundary condition, a choice of polarization on the topological face, and the degrees of freedom living on the corners (which we are going to refer as corner theory) of the boundary SymTFT.

For the compact boson at radius $R$ the SymTFT is the three-dimensional BF theory of two $\mathbb{R}$-valued gauge fields, whose line operators generate the momentum and winding $U(1)$ symmetries. Moreover, the braiding between the line operators in the bulk of the SymTFT encodes the mixed anomaly of the free boson theory \cite{Antinucci:2024zjp, Argurio:2024ewp, Robbins:2025urk}. A feature of the SymTFT formulation of the free boson that will be important for us is that the information of the radius of the free boson enters only through the coupling of an edge mode on the symmetry boundary, while the physical boundary condition remains the same for every $R$. We show that the two natural topological conditions on the topological face of the boundary SymTFT, together with the corner modes, reproduce exactly edge mode theories required to impose the Dirichlet and the Neuman boundary conditions. 

We then study how T-duality acts within this boundary SymTFT construction. We first review the codimension one surface defect $\mathcal T_s$ of the free boson SymTFT, following \cite{Argurio:2024ewp}, which exchanges the bulk fields, and maps the symmetry boundary at radius $R$ to the symmetry boundary at radius $s/R$. We subsequently place this defect vertically in the boundary SymTFT,  fusion with the symmetry boundary dualizes its edge mode localized on the symmetry boundary. Consequently, For $s=1$, the Dirichlet corner modes at radius $R$ is thereby mapped to the Neumann corner modes at radius $1/R$. In this way, we show that the T-duality exchanges the corner degrees of freedom. In particular, we show how T-duality exchanges the Dirichlet and Neumann boundary conditions of the free boson theory.
\begin{equation}
    \mathcal T_1[D]\otimes\mathsf D_R=\mathsf N_{1/R}
\end{equation}

We also place the defect horizontally across the slab, the defect becomes the T-duality line of the two dimensional theory \cite{Argurio:2024ewp}. At the self-dual radius both of its ends are transparent, the line is an invertible symmetry, and fusing it with the face exchanges the two corner theories at fixed radius. Away from the self-dual radius its end on the symmetry boundary is a genuine junction, which only admits some of the winding sectors of the edge modes; this is how the non-invertibility of T-duality at a generic radius shows up in the boundary SymTFT.

The novelty is therefore not in the identification of the familiar Dirichlet and Neumann conditions, rather, the explicit realization of the Dirichlet and Neumann boundary conditions of the compact boson using topological edge modes. To the best of our knowledge, this has not been done before. Moreover, the identification of the edge degrees of freedom with the degrees of freedom localized on the corners of the boundary SymTFT, together with a choice of polarization on the topological face.

The final sections of the paper consider a couple of generalizations to the topological edge mode approach.  First, we consider the possibility of including non-compact scalar edge modes.  One motivation is that this allows for the possibility of a continuous spectrum of boundary operators, which would be a prerequisite for describing Friedan-Janik (FJ) boundaries~\cite{Friedan:99, Janik:2001hb}.  Such a description, via an explicit action, would be interesting for study of these boundaries.  In particular, it might provide a better handle on understanding deformations of such non-standard boundaries.  In fact we are able to get continuous spectra and we argue that we can realize smeared Dirichlet and Neumann boundaries, which can be thought of as limiting cases of FJ boundaries, by including a single non-compact edge mode.  However the full set of FJ boundaries does not seem to be amenable to these techniques.

Secondly, we briefly consider the generalization of having multiple compact free bosons.  We show that one can recover the standard boundary conditions with Dirichlet in some directions and (fluxed) Neumann in other directions.

This paper is organized as follows. In Section \ref{sec: free boson boundary}, we introduce the simplest edge mode actions that impose Dirichlet and Neumann boundary conditions, and then we consider a general action with $N$ edge modes. We conclude this section by calculating the OPEs of the stress tensor with the vertex operators made out of the edge modes. In Section \ref{sec: boundary symtft}, the boundary SymTFT analysis for the compact free boson with a boundary is performed, and the T-duality defect is studied. In Section \ref{sec: Noncompact edge modes}, we consider the non-compact edge modes and show that they produce smeared Dirichlet and Neumann states, which correspond to the edge cases for the Friedan-Janik states. In Section \ref{sec: multiple bulk}, we consider a generalization with multiple bulk fields, and show that we can infer the number of Dirichlet branes from the data appearing in the edge mode action. Finally, we have provided some details of the calculations used in the main text, in appendices~\ref{app: Path integral} and~\ref{app: appendix 1}.

\section{Compact free boson on a manifold with boundary} \label{sec: free boson boundary}

The action of the compact free boson CFT is
\be\label{free boson bulk action}
S_{\text{free boson}} = \frac{R^2}{4\pi} \int_{\mathcal{M}_2} dX \wedge * dX,
\ee
where $\al'$ is set to $1$, and we will use this value throughout this paper. Moreover, the T-dual of $X$ will be denoted as $\wt{X}$. We have chosen the normalization so that the period of $X$ is $2\pi$.  This implies that $\wt{X}$ has period $2\pi/R^2$. Lastly, in Euclidean signature, we have $*dX=-id\wt{X}$. 

The variation of \eqref{free boson bulk action} is\footnote{Throughout this section, all bulk differential forms appearing in boundary integrals are implicitly understood to be pulled back along the inclusion $\iota: \partial\mathcal{M}_2 \hookrightarrow \mathcal{M}_2$. Thus, $dX$ inside boundary integral denotes $\iota^* (dX)$. We suppress the pullback map ($\iota^*$) to avoid clutter.}
\begin{equation}\label{bulk action variation}
\begin{aligned}
      \delta S &= \frac{R^2}{2\pi} \int_{\mathcal{M}_2} d(\delta X)  *dX \\
      &= - \frac{R^2}{2\pi} \int_{\mathcal{M}_2} \delta X  d * dX + \frac{R^2}{2\pi} \int_{\partial \mathcal{M}_2} \delta X *dX. 
\end{aligned}
\end{equation}
Therefore, the bulk equation of motion is $d \star dX =0$, and the remaining boundary variation can vanish if we impose certain boundary conditions. There are two such canonical boundary conditions that cancel the boundary variation, i.e., Dirichlet and Neumann boundary conditions which are given as 
\begin{equation} \label{eq: D and N}
    \text{Dirichlet}: \d X = 0 , \qquad \text{Neumann}:~ *d X =0.
\end{equation}
Next, we would like to reproduce these boundary conditions by adding edge degrees of freedom ($U(1)$ valued $2\pi$-periodic scalars) on the boundary. 

\subsection{Imposing boundary conditions using edge modes}

We can also impose the boundary conditions in \eqref{eq: D and N} using a boundary TFT (in this case, it will be a one-dimensional TFT). Consider the following TFT,
\begin{equation}\label{eqn:dirichlet_TFT}
S^{\text{Dirichlet}}_{\text{edge mode}} : = \frac{i}{2\pi}\int_{\partial \mathcal{M}_2} \Psi \wedge dX,
\end{equation}
where $\Psi$ is a $U(1)$-valued scalar localized on the boundary\footnote{This is very similar to coupling a QFT to a topological degrees of freedom in the spirit of \cite{Kapustin:2014gua} but this time the topological degrees of freedom are localized on to the boundary only.}. The combined boundary variation gives 
\begin{equation}
\begin{aligned}
    \delta ( S_{\text{free boson}} + S^{\text{Dirichlet}}_{\text{edge mode}} )  &= \text{bulk} +   \frac{R^2}{2\pi} \int_{\partial \mathcal{M}_2} \delta X  \star dX + \frac{i}{2\pi} \int_{\partial \mathcal{M}_2} \left(\delta \Psi dX - d\Psi \wedge \delta X\right) \\
    &= \text{bulk} +  \frac{1}{2\pi} \int_{\partial \mathcal{M}_2}  \left[\delta X (R^2\star dX - id\Psi) +  i\delta \Psi  dX \right],
\end{aligned}    
\end{equation}
and the boundary equations of motion coming from this variation enforce
\begin{equation}
\Psi: ~ dX =0 , \qquad X: ~ R^2 \star dX = id\Psi.
\end{equation}
This is the Dirichlet boundary condition on $X$, since $dX=0$ is the derivative of $X$ along the boundary. In Appendix \ref{app: Path integral}, we argue that using the edge mode action \eqref{eqn:dirichlet_TFT}, the resulting density of states matches the density of states expected from a simple Dirichlet state. We can also impose Neumann boundary conditions by introducing an additional boundary degree of freedom as
\begin{equation}\label{eqn:Neumann_TFT}
S^{\text{Neumann}}_{\text{edge mode}} : = \frac{i}{2\pi} \int_{\partial  \mathcal{M}_2} \Big(\Psi_1 \wedge d\Psi_2 + \Psi_2 \wedge dX \Big), 
\end{equation}
where, as before, $\Psi_i$ are $U(1)$-valued fields localized on the boundary. The (boundary piece of the) variation of the full action gives
\begin{equation}
\begin{aligned}
    &\delta (S_{\text{free boson}} + S^{\text{Neumann}}_{\text{edge mode}}) \\
    &= \frac{R^2}{2\pi} \int_{\partial \mathcal{M}_2}\delta X \star dX + \frac{i}{2\pi} \int_{\partial \mathcal{M}_2} \left[\delta \Psi_1  d\Psi_2 + \Psi_1 d\delta \Psi_2 
    + \delta \Psi_2  dX + \Psi_2  d\delta X\right].
\end{aligned}
\end{equation}
We can rewrite this by doing a couple of integrations by parts, 
\begin{equation}
    \delta (S_{\text{free boson}} + S^{\text{Neumann}}_{\text{edge mode}}) = \frac{1}{2\pi} \int_{\partial \mathcal{M}_2} \left[\delta X  \left( R^2 *dX - id\Psi_2 \right) + i\delta \Psi_1  d\Psi_2 - i\delta \Psi_2  \left( d\Psi_1 - dX\right)\right].
\end{equation}
The boundary equations of motion enforce the following conditions, 
\begin{equation}\label{eq: Neumann-edge-mode-equations}
\delta \Psi_1:  d  \Psi_2 =0, \qquad \delta X :  R^2 \star dX - id\Psi_2 =0, \qquad\delta \Psi_2:  dX = d\Psi_1.
\end{equation}
The first and second equations imply $R^2 \star dX =0$, and it allows us to identify this boundary condition with the Neumann boundary condition.

\subsubsection{Family of Dirichlet and Neumann boundary conditions}

We have identified the boundary TFT that enforces the canonical boundary conditions, but it is well known that free boson theory allows a family of Dirichlet and Neumann boundary conditions (see, e.g., \cite{Oshikawa:1996}) labeled by the zero mode of $X$ and $\wt{X}$ respectively. In order to generate the family of Dirichlet boundary conditions, we modify the edge mode TFT \eqref{eqn:dirichlet_TFT} as, 
\begin{equation}\label{Dirichlet_Family_TFT}
\boxed{S^{\text{D}}_{\text{edge mode}} = \frac{i}{2\pi}\int_{\partial \mathcal{M}_2} \Big( \Psi \wedge dX + \al d\Psi \Big),}
\end{equation}
where $\al$ is a $2\pi$ periodic real parameter on the boundary. We can think of the last term as the insertion of a $\theta$-angle. Insertion of such terms does not modify the boundary equation of motion, but since $X$ and $\Psi$ are compact fields, summing over winding sectors imposes quantization conditions on the coefficients of $dX$ and $d\Psi$ in the boundary action. We will refer to these constraints as quantization constraints. The $\Psi$ quantization constraint implies
\begin{equation}\label{eq: Dirichlet family generation}
    X - \al \in 2\pi\Z.
\end{equation}
Since the value of $X$ is fixed to $\al$ (mod $2\pi$), different values of $\al$ label the members of the family of Dirichlet boundaries.  Therefore, the boundary edge mode TFT generates the family of Dirichlet boundary conditions parametrized by $\al$.

Similarly, we can also reproduce the family of Neumann boundary conditions by adding a $\theta$-angle-like term to \eqref{eqn:Neumann_TFT}, which will not modify the boundary equations of motion as
\begin{equation}
\boxed{S^{\text{N}}_{\text{edge mode}} = \frac{i}{2\pi}\int_{\partial \mathcal{M}_2} \Big( \Psi_1 \wedge d\Psi_2 + \Psi_2 \wedge dX + \al d\Psi_1 \Big).}
\end{equation}
quantization constraints from $\Psi_1, \Psi_2$ and $X$ imply
\begin{align}
    \Psi_2-\al\in\ & 2\pi\mathbb{Z},\\
    \Psi_1-X\in\ & 2\pi\mathbb{Z},\\
    R^2\widetilde{X} +\Psi_2\in\ & 2\pi\mathbb{Z}.
\end{align}
The first and third equations collectively imply that the value of $\wt{X}$ is fixed to be
\be
\wt{X}=-\frac{\al}{R^2}\text{ mod }\frac{2\pi}{R^2},
\ee
which is exactly what we expect for a Neumann boundary.

\subsection{A generalization of the boundary theory}
Now we are going to generalize the edge-mode theory and analyze the boundary conditions. The most general boundary theory that we will consider contains $N$ edge modes $\Psi_i$ and is given as
\be
     S^{\text{General}}_{\text{Edge}} = \frac{i}{2\pi} \int_{\p \M}\lp \sum_{i=1}^{N}v_{i}\Psi_{i} dX + \sum_{i,j=1}^{N}\frac{k_{ij}}{2}\Psi_{i}\;d\Psi_{j}+\al_0 dX+ \sum_{i=1}^{N}\al_{i}\;d\Psi_{i} \rp,
\ee
where $v_{i}$ is an $N$ dimensional vector of integers and $k_{ij}$ is an antisymmetric $N\times N$ matrix of integers. We can think of the last term as one-dimensional $\theta$-angles. We can do a field redefinition of the edge modes from $\Psi_i$ to $M_{ij}\Psi_j$ where $M\in GL(N,\Z)$. This transformation can be seen instead as acting on $v_i$ and $k_{ij}$ as
$$
v_i\rr(M^Tv)_i,\efill k_{ij}\rr(M^TkM)_{ij},
$$
and we can choose an $M$ such all $v_i$'s except $v_1$ are set to zero\footnote{This unfortunately is a different convention than we used for the two-mode action~\eqref{eqn:Neumann_TFT} encoding Neumann boundary conditions, in which $X$ only coupled directly to $\Psi_2$.} and $k_{ij}$ is set to the skew-tridiagonal form i.e.
$$
\begin{pmatrix}
    0 & k_1 & 0 & \cdots\\
    -k_1 & 0 & k_2 & \cdots\\
    0& -k_2 & 0&\cdots\\
    \vdots&\vdots&\vdots&\ddots
\end{pmatrix}.
$$
Setting the new $v_1$ as $v$, the full action becomes the following
\begin{equation}\label{general lagrangian}
S^{\text{General}}_{\text{Edge}} =\frac{R^2}{4\pi}\int_M dX\wedge *dX+\frac{i}{2\pi}\int_{\p M}\lp v\Psi_1 dX+\sum_{i=1}^{N-1}k_i\Psi_id\Psi_{i+1}+\al_0dX+\sum_{i=1}^{N}\al_id\Psi_i\rp. 
\end{equation}
Note that if $v$ or any of the $k_i$ vanish, then some of the $\Psi_i$ are completely decoupled from the bulk theory.  In this case, we are simply stacking a decoupled TFT at the boundary.  This case is not actually interesting, so we shall assume that the integers $v$ and $k_i$ are all non-zero.

The equations of motion at the boundary for $X, \Psi_1, \Psi_i\;(1<i<N)$, and $\Psi_N$ respectively are given as
\begin{align}\label{compact edge modes eom X}
R^2*dX- ivd\Psi_1=0,\\
\label{compact edge modes eom psi1}
vdX+k_1d\Psi_2=0,\\
\label{compact edge modes eom psii}
k_id\Psi_{i+1}-k_{i-1}d\Psi_{i-1}=0,\\
\label{compact edge modes eom psiN}
k_{N-1}d\Psi_{N-1}=0.
\end{align}
Moreover, the quantization constraints of these fields gives the following conditions, respectively
\begin{align}
R^2\tilde{X}+v\Psi_1+\al_0\in2\pi\Z,\\
-vX-k_1\Psi_2+\al_1\in2\pi\Z,\\
k_{i-1}\Psi_{i-1}-k_i\Psi_{i+1}+\al_i\in2\pi\Z,\\
k_{N-1}\Psi_{N-1}+\al_N\in2\pi\Z,
\end{align}
which allows us to write everything in terms of $\Psi_1$ and $\Psi_2$
\begin{align}\label{gen case winding 1}
X=-\frac{k_1}{v}\Psi_2+\frac{\al_1}{v}\text{ mod }\frac{2\pi \Z}{v},\\
\label{gen case winding 2}
\tilde{X}=-\frac{v}{R^2}\Psi_1-\frac{\al_0}{R^2}\text{ mod }\frac{2\pi \Z}{R^2},\\
\label{gen case winding 3}
\Psi_{i+1}=\frac{k_{i-1}}{k_i}\Psi_{i-1}+\frac{\al_i}{k_i}\text{ mod }\frac{2\pi \Z}{k_i},\\
\label{gen case winding 4}
\Psi_{N-1}=-\frac{\al_{N}}{k_{N-1}}\text{ mod }\frac{2\pi \Z}{k_{N-1}}.
\end{align}
We can see that $\Psi_{N-1}$ is a constant, and if $N$ is even (odd), all $\Psi_{i}$ with odd (even) $i$'s are constants and $d\tilde{X}$ ($dX$) vanishes, which leads us to Neumann (Dirichlet) boundary. This is consistent with our previous finding that having one edge mode led to the family of Dirichlet boundaries, while having two edge modes led to the Neumann family. If any given $k_j$ vanishes, it renders all succeeding $\Psi$'s of the same parity as $j$ constant (i.e. all $\Psi_{j+2q}$ for $q\in\Z^{+}$ are constant.)

We can solve \eqref{gen case winding 1}-\eqref{gen case winding 4} to write all the edge modes in terms of $X$ and $\tilde{X}$. The relevant results that will be useful to calculate the OPEs in subsection \ref{operator algebras} are
\begin{align}\label{even psi in terms of X}
\Psi_{2j}&=-\frac{v}{k_1}\lp\prod_{r=1}^{j-1}\frac{k_{2r}}{k_{2r+1}}\rp X+j\text{ dependent constant}\;\;\;\;(j>1),\\
\label{odd psi in terms of X}
\Psi_{2j+1}&=-\frac{R^2}{v}\lp\prod_{r=1}^{j}\frac{k_{2r-1}}{k_{2r}}\rp\wt{X}+j\text{ dependent constant}\;\;\;\;(j>0).
\end{align}
For the $j=1$ case in \eqref{even psi in terms of X} and the $j=0$ case in \eqref{odd psi in terms of X}, the product in the parentheses should be excluded.

\subsection{The operator algebras}\label{operator algebras}
We will now show that the spectra of the conformal weights of $\Psi_i$s that we get in the Dirichlet and Neumann cases match the corresponding spectra of boundary operators. To do this, we will compute OPEs of the vertex operators made out of $\Psi_i$s with the holomorphic stress tensor. Let's discuss some preliminary results first. The usual bulk-bulk OPE for the compact free boson is (recall that we normalize the periodicity of $X$ to $2\pi$ and set $\al'=1$)
\be
X(z,\bz)X(w,\bar{w})\sim -\frac{1}{2R^2} \ln|z-w|^2,
\ee
and the holomorphic stress tensor is
\be
T(z)=-R^2(\p X)^2(z).
\ee
In the presence of a boundary, we can use the method of images \cite{Cardy:1984bb} to have the holomorphic field $X_L(z)$ in the upper half-plane, the anti-holomorphic field $X_R(\bz)$ in the lower half-plane, with these two fields identified on the boundary (where $z=\bz$) via some gluing map depending on the boundary condition. For Dirichlet and Neumann boundaries, the gluing maps imply (these follow from $X=\text{constant}$ and $\iota^\ast(\ast dX)=0$ respectively, with $X(z,\bar{z})=X_L(z)+X_R(\bar{z})$)
\begin{align}
X_L\vert_{z=\bz}=-X_R\vert_{z=\bz}+\text{constant}\efill (\text{Dirichlet}),\\
X_L\vert_{z=\bz}=X_R\vert_{z=\bz}+\text{constant}\efill (\text{Neumann}).
\end{align}
Therefore, we have the following results for the Dirichlet boundary (with $x$ on the boundary)
\begin{align}
X_L(z)X_R(x)=-X_L(z)X_L(x)\sim\frac{1}{2R^2} \ln|z-x|^2,\\
X_R(\bz)X_L(x)=-X_R(\bz)X_R(x)\sim\frac{1}{2R^2} \ln|\bz-x|^2,
\end{align}
and the following results for the Neumann boundary
\begin{align}
X_L(z)X_R(x)=X_L(z)X_L(x)\sim-\frac{1}{2R^2} \ln|z-x|^2,\\
X_R(\bz)X_L(x)=X_R(\bz)X_R(x)\sim-\frac{1}{2R^2} \ln|\bz-x|^2.
\end{align}
Using these results, we have the following OPE results
\begin{align}
T(z):e^{ik \wt{X}(x)}:\;\sim \frac{k^2}{R^2(z-x)^2}e^{ik \wt{X}(x)}\;\;\;\lp\text{Dirichlet}\rp,\\
T(z):e^{ik X(x)}:\;\sim \frac{k^2}{R^2(z-x)^2}e^{ik X(x)}\;\;\;\lp\text{Neumann}\rp,
\end{align}
for arbitrary $k$ and where the dual scalar $\wt{X}$ is defined as $\wt{X}=X_L-X_R$.

Now, for a single edge mode, if we consider the vertex operator $\mathcal{V}_{n}(x)=:e^{in\Psi(x)}:$, with $n\in \Z$, and calculate its OPE with $T(z)$, using the fact that the quantization constraints imply
\be
\Psi(x)=-\frac{R^2}{v}\widetilde{X}(x)-\frac{\al_0}{v}-\frac{2\pi\Z}{v},
\ee
we get
\be\label{T ope with e^inpsi}
T(z)\mathcal{V}_{n}(x)\sim \frac{n^2R^2}{v^2}\frac{\mathcal{V}_{n}(x)}{(z-x)^2}+\frac{\p\mathcal{V}_{n}(x)}{z-x}.
\ee
Now, in the boundary state formalism \cite{Cardy:1984bb}, the boundary states for Dirichlet and Neumann boundaries are written as \cite{Oshikawa:1996}
\begin{align}
    ||D(x_0)\rrangle&=\frac{1}{\sqrt{R\sqrt{2}}}\sum_{N\in\Z}e^{-iNx_0}||(N,0)\rrangle,\\
    ||N(\wt{x}_0)\rrangle&=\sqrt{\frac{R}{\sqrt{2}}}\sum_{M\in\Z}e^{-iMR^2\wt{x}_0}||(0,M)\rrangle,
\end{align}
where $||(N,M)\rrangle$ are the Ishibashi states corresponding to the primary state with momentum $N$ and winding $M$. The parameters $x_0$ and $\wt{x}_0$ have periods $2\pi$ and $2\pi/R^2$ respectively. The overlap of these boundaries with themselves is (e.g., see \cite{Oshikawa:1996})
\begin{align}\label{dirichet overlap}
\llangle D(x_0)| q^H | D(x_0')\rrangle&=\sum_{n\in\Z}\chi^{U(1)}_{R^2\lp n-\frac{x_{0}-x'_0}{2\pi}\rp^2}(\widetilde{q}),\\
\label{neumann overlap}
\llangle N(\wt{x}_0)| q^H | N(\wt{x}_0')\rrangle&=\sum_{m\in\Z}\chi^{U(1)}_{\frac{1}{R^2}\lp m+\frac{R^2}{2\pi}(\wt{x}_{0}-\wt{x}'_0)\rp^2}(\widetilde{q}),
\end{align}
where $q=e^{2\pi i\tau}$ with $\tau$ being the torus modulus. The conformal weights $h$ appearing in the characters $\chi^{U(1)}_{h}(\wt{q})$ in \eqref{dirichet overlap} are the conformal weights of the boundary fields for $x_0=x'_0$. These weights match the $n^2R^2$ weights appearing in the spectrum of $\Psi$ for $|v|=1$ in \eqref{T ope with e^inpsi}. For $|v|>1$, we can consider superpositions of Dirichlet boundaries, which correspond to $|v|$ equidistant $D0$ branes on the target space circle. These superpositions have the following form
\be\label{Dirichlet superposition}
||\mathcal{D}(\overrightarrow{a})\rrangle=\sum_{\ell=0}^{|v|-1}a_k \left|\left|D\lp \frac{2\pi \ell}{|v|}\rp\right.\right\rrangle\;\;\;\text{ where }a_\ell\in\Z_{\geq 0}.
\ee
The overlap of this superposition with itself is
\begin{align}
\llangle \mathcal{D}(\overrightarrow{a})| q^H|\mathcal{D}(\overrightarrow{a})\rrangle&=\sum_{\ell,\ell'=0}^{|v|-1}a_\ell a_{\ell'}\left\llangle D\lp\frac{2\pi \ell}{|v|}\rp\;\Big\lvert\;q^H\;\Big\rvert\;D\lp \frac{2\pi \ell'}{|v|}\rp\right\rrangle\nonumber\\
&=\sum_{\ell,\ell'=0}^{|v|-1}a_\ell a_{\ell'}\sum_{j\in\Z}\chi^{U(1)}_{R^2\lp j-\frac{\ell-\ell'}{|v|}\rp^2}(\widetilde{q}).
\end{align}
One can see that the combination $j|v|-(\ell-\ell')$ with $0\leq \ell,\ell'\leq |v|-1$ and $j\in \Z$ sweeps through all integers and thus can be replaced by an arbitrary integer $n$. Therefore, the conformal weights of the boundary operators for the boundary state \eqref{Dirichlet superposition} are just $R^2n^2/v^2$, which match the conformal weights derived from the OPE \eqref{T ope with e^inpsi}. Hence, the action \eqref{general lagrangian} with $N=1$ and $|v|\geq 1$  may correspond to a set of $|v|$ $D0$ branes at angles $2\pi \ell/|v|$ with $0\leq \ell\leq |v|-1$ on the target space circle.

For the case of two edge modes, i.e., $N=2$, we can take a generic vertex operator as $\mathcal{V}_{n,m}(x)=:e^{in\Psi_1(x)+im\Psi_2(x)}:$
and using the quantization constraints, we get the following OPE
\be\label{T OPE with e^in Psi1 +im Psi2}
T(z) \mathcal{V}_{n,m}(x) \sim \frac{m^2v^2}{k^2R^2}\frac{\mathcal{V}_{n,m}(x)}{(z-x)^2}+\frac{\p_t\mathcal{V}_{n,m}(x)}{z-x}.
\ee
The fact that the conformal weight depends only on $m$ and not on $n$ is simply a consequence of the fact that the equations of motion fix $\Psi_1$ to be a constant, so $e^{in\Psi_1(x)}$ is simply an inessential constant phase.

Now, we consider a superposition of Neumann states which corresponds to $|k|$ equidistant $D0$ branes on the dual target space circle
\be\label{Neumann superposition}
||\mathcal{N}(\overrightarrow{a})\rrangle=\sum_{\ell=0}^{|k|-1}a_\ell \left|\left|N\lp \frac{2\pi \ell}{|k|R^2}\rp\right.\right\rrangle\;\;\;\text{ where }a_\ell\in\Z_{\geq 0},
\ee
which gives the following overlap
\be
\llangle \mathcal{N}(\overrightarrow{a})| q^H|\mathcal{N}(\overrightarrow{a})\rrangle=\sum_{\ell,\ell'=0}^{|k|-1}a_\ell a_{\ell'}\sum_{j\in\Z}\chi^{U(1)}_{\frac{1}{R^2}\lp j+\frac{\ell-\ell'}{|k|}\rp^2}(\widetilde{q}),
\ee
leading to conformal weights of the form $m^2/k^2R^2$. These conformal weights match those in \eqref{T OPE with e^in Psi1 +im Psi2} for $|v|=1$. It is not clear what should be the interpretation of the $|v|>1$ cases here.

We can calculate the OPE of the stress tensor with a general vertex operator for $N$ edge modes using \eqref{even psi in terms of X} and \eqref{odd psi in terms of X}. For even (odd) $N$, the non-trivial part of the vertex operators is made of $\Psi_i$ with even (odd) $i$ as the other $\Psi_i$'s are constant and just provide a phase. For even and odd $N$, we can define general vertex operators as
\begin{align}
\mathcal{V}_{N,\overrightarrow{m}}&=\;(\text{Phase}):\exp\lp i\sum_{j=1}^{N/2}m_{j}\Psi_{2j}\rp:\;\;\;\;(\text{even }N),\\
\mathcal{V}_{N,\overrightarrow{m}}&=\;(\text{Phase}):\exp\lp i\sum_{j=1}^{\frac{N+1}{2}}m_{j}\Psi_{2j-1}\rp:\;\;\;\;(\text{odd }N),
\end{align}
with $m_j\in \Z$ and using \eqref{even psi in terms of X}, we have the following results
\begin{align}
T(z)\mathcal{V}_{N,\overrightarrow{m}}(x)&\sim\frac{v^2}{R^2k^2_1}\ls m_1+\sum_{j=2}^{N/2}m_j\lp\prod^{j-1}_{r=1}\frac{k_{2r}}{k_{2r+1}} \rp\rs^2\frac{\mathcal{V}_{N,\overrightarrow{m}}(x)}{(z-x)^2}+\cdots\;\;\;\;(\text{even }N),\\
T(z)\mathcal{V}_{N,\overrightarrow{m}}(x)&\sim\frac{R^2}{v^2}\ls m_1+\sum_{j=2}^{\frac{N+1}{2}}m_j\lp\prod^{j}_{r=1}\frac{k_{2r-1}}{k_{2r}} \rp\rs^2\frac{\mathcal{V}_{N,\overrightarrow{m}}(x)}{(z-x)^2}+\cdots\;\;\;\;(\text{odd }N).
\end{align}
These expressions agree with \eqref{T ope with e^inpsi} and \eqref{T OPE with e^in Psi1 +im Psi2} for $N=1$ and $N=2$. We can see that for arbitrary $N$, the spectrum of conformal weights is still discrete. Therefore, we can't get a continuous spectrum of conformal weights using just the compact edge modes. We will revisit this point in section \ref{sec: Noncompact edge modes}.

\section{Boundary SymTFT and T-duality}\label{sec: boundary symtft}

In the previous section, we constructed topological edge theories that implement Dirichlet and Neumann boundary conditions for the compact free boson directly in two dimensions. In this section, we show that these edge theories admit a natural higher-dimensional realization: they arise as corner theories of the compact-boson SymTFT. Recall that the momentum and winding symmetries of the compact boson form a ($U(1)\times U(1)$) symmetry with a mixed anomaly, captured by a three-dimensional noncompact BF theory \cite{Antinucci:2024zjp,Brennan:2024fgj,Argurio:2024ewp}. We begin by reviewing its SymTFT and verifying that interval compactification reproduces the compact boson at radius $R$. The main new ingredient appears when the physical spacetime $\Sigma_2$ itself has a boundary. The SymTFT manifold $M_3=\Sigma_2\times[0,1]$ then acquires an additional topological face, which meets the symmetry and physical boundaries along codimension-two corners. 

\subsection{SymTFT for the compact free boson}
We are going to review the SymTFT analysis of the compact free boson theory. Primarily, we are going to follow \cite{Argurio:2024ewp}. The three-dimensional SymTFT capturing the $U(1) \times U(1)$ symmetry with a mixed anomaly of the compact free boson theory is given by \cite{Antinucci:2024zjp,Brennan:2024fgj},
\begin{equation}
\boxed{    S_{\text{SymTFT}} = \frac{i}{2\pi} \int_{M_3} a \wedge db,}
\end{equation}
where both $a$ and $b$ are $\mathbb{R}$-valued gauge fields with the following gauge transformations.
\begin{equation}
\begin{aligned}
    a &\rightarrow a + d\lambda_a, \\ 
    b &\rightarrow b + d\lambda_b. 
\end{aligned}    
\end{equation}
The topological operators in the bulk are given by,  
\begin{equation}
    U_m[\gamma] = \mathrm{exp} \,\Bigl( im\int_{\gamma} a \Bigr), \qquad V_n[\delta] = \mathrm{exp}\, \Bigl( in\int_{\delta} b \Bigr), 
\end{equation}
where $m,n \in \mathbb{R}$. The mutual braiding between the lines is given by, 
\begin{equation}
    \langle U_m[\gamma], V_n[\delta] \rangle = \text{exp} \, \bigl( 2\pi i\, mn \,\text{Link}(\gamma, \delta) \bigr)\,.
\end{equation}
This non-trivial braiding is the SymTFT manifestation of the mixed anomaly between momentum and winding $U(1)$ symmetries \cite{Antinucci:2024zjp}.

\subsubsection*{The boundaries and the interval compactification}

In order to obtain the two dimensional theory, we need to specify the boundary conditions of the SymTFT and then close the sandwich. We place the SymTFT in $M_3 := \Sigma_2 \times [0,1]$, with the topological symmetry boundary at $\Sigma_2 \times \{0\}$ and the non-topological physical boundary at $\Sigma_2 \times \{1\}$. After the interval compactification, we are going to recover the two-dimensional compact free boson theory on $\Sigma_2$.

All the integrals over the two ends of the interval are taken with orientation of $\Sigma_2$, and $M_3$ is oriented such that the Stokes theorem reads as, 
\begin{equation}\label{eq: orientation}
    \int_{M_3} d\omega = \int_{\Sigma_2 \times \{ 1\}} \omega - \int_{\Sigma_2 \times \{ 0\}} \omega,
\end{equation}
i.e.~$\partial M_3 = \Sigma_2 \cup \bar{\Sigma}_2$ in the notation of \cite{Argurio:2024ewp}, the symmetry boundary being the component with
reversed orientation.

The topological symmetry boundary is given by 
\begin{equation}\label{eq: sym-boundary-edge}
    S_{\text{Sym Bdry}} = - \frac{iR}{2\pi} \int_{\Sigma_2 \times \{0\} } X \wedge db,
\end{equation}
where we have introduced an edge mode $X$ on the symmetry boundary, which is a $U(1)$-valued $2\pi$-periodic scalar ($X \sim X + 2\pi$), with the following gauge transformation, 
\begin{equation}
    X \rightarrow X - R^{-1} \lambda_a. 
\end{equation}
The overall sign in \eqref{eq: sym-boundary-edge} accounts for the orientation of the symmetry boundary in \eqref{eq: orientation}. Moreover, this boundary action is designed to cancel the gauge non-invariance of the bulk action. Under $a \rightarrow a + d\lambda_a$,
\begin{equation}
    \Delta S_{\text{SymTFT}} = \frac{i}{2\pi} \int_{M_3} d\lambda_a db = \frac{i}{2\pi}  \int_{\Sigma_2 \times \{1\}} \lambda_a db -    
    \frac{i}{2\pi} \int_{\Sigma_2 \times \{0\}} \lambda_a db.
\end{equation}
The second term is canceled by the variation of \eqref{eq: sym-boundary-edge} under $X \rightarrow X - R^{-1} \lambda_a $. On the physical boundary the gauge transformations of $a$ and $b$ must be restricted to be compatible with the non-topological boundary condition introduced below. This information will play a crucial role when we eventually introduce corners to deal with boundaries. 

We write the bulk--boundary combined action as,
\begin{equation}
    S_{\text{comb}} = \frac{i}{2\pi} \int_{M_3} a \wedge db - \frac{iR}{2\pi} \int_{\Sigma_2 \times \{0\}} X \wedge db,
\end{equation}
and then we perform a classical variation, 
\begin{equation}
\begin{aligned}
    \delta S_{\text{comb}} &= \frac{i}{2\pi} \int_{M_3} \left(\delta adb + ad\delta b\right) - \frac{iR}{2\pi} \int_{\Sigma_2 \times \{0\}} \Big( \delta X db + X d\delta b \Big) \\
    &= \frac{i}{2\pi} \int_{M_3} \left(\delta adb + da\delta b\right) -\frac{i}{2\pi} \int_{\Sigma_2 \times \{ 1\}} a\delta b
    + \frac{i}{2\pi}\int_{\Sigma_2 \times \{0\}} \left[(a + RdX )\delta b - R \delta X db\right].
\end{aligned}
\end{equation}
The bulk equations of motion are $da=db=0$ and the boundary equations on the symmetry boundary tell us, 
\begin{equation}
    a \big|_{\Sigma_2 \times \{0\}} = - R ~dX, \qquad db\big|_{\Sigma_2 \times \{0\} } =0.
\end{equation}
\begin{figure}[ht]
    \centering
    \begin{tikzpicture}[scale = 1.5]
        \fill[blue!5] (0,0) rectangle (7,2.7);

        \draw[line width=2.2pt, blue!65!black]    (0,0) -- (0,2.7);

        \draw[line width=2.2pt, red!65!black] (7,0) -- (7,2.7);

        \node[align=center, anchor=south] at (0,2.85)  {$\mathcal{B}_{\mathrm{Sym}} =  -\frac{iR}{2\pi} \int X \wedge db$};

        \node[align=center, anchor=south] at (7,2.85){$\mathcal{B}_{\mathrm{Phys}} = \frac{1}{4\pi} \int  b \wedge \star b$};

        \node[align=center]  at (3.5,1.55) {$ S_{\text{SymTFT}} = \frac{i}{2\pi} \int_{M_3} a \wedge db$ };

        \node[align=center]  at (3.5,0.25) {$M_3=\Sigma_2\times [0,1]$};

    \end{tikzpicture}

    \caption{The SymTFT sandwich for the compact free boson. The metric dependent physical boundary and the topological symmetry boundary are placed at the two ends of the interval.}
    \label{fig:compact-boson-symtft}
\end{figure}

On the other hand, we choose the following as our non-topological physical boundary (see Figure~\ref{fig:compact-boson-symtft}), 
\begin{equation}
    S_{\text{Phys Bdry}} = \frac{1}{4\pi} \int_{\Sigma_2 \times \{1\}}  b \wedge \star b,
\end{equation}
where we have introduced a Hodge star (and hence a background metric) on $\Sigma_2 \times \{ 1\}$. Its variation, combined with the boundary contribution picked up from the variation of the bulk term is given by,
\begin{equation}
    \begin{aligned}
        \delta S = \text{bulk terms} + \frac{1}{2\pi} \int_{\Sigma_2 \times \{1\}} \delta b \bigl( \star b +ia \bigr),
    \end{aligned}
\end{equation}
so, the boundary equation of motion enforce, 
\begin{equation}
    a\big|_{\Sigma_2 \times \{1\}} = i \star b  \;\;\Rightarrow b\big|_{\Sigma_2 \times \{1\}} = i*a,
\end{equation}
(recall that $\ast^2=-1$ for one-forms in two Euclidean dimensions). Note that, to make sure the action on the physical boundary is gauge invariant, we have to restrict the gauge transformations of bulk field $b$, on the physical boundary \cite{Banerjee:2026wop}. 

Now, we perform interval compactification. More precisely, we integrate out the three dimensional bulk degrees of freedom while keeping the boundary scalar $X$. 

At the level of local differential forms, the bulk BF term may be integrated by parts using our orientation convention, 
\begin{equation}
    \frac{i}{2\pi} \int_{M_3} a\, db = \frac{i}{2\pi} \int_{M_3} da\, b - \frac{i}{2\pi} \int_{\Sigma_2\times \{1\}} a\, b + \frac{i}{2\pi} \int_{\Sigma_2 \times\{0\}} a\, b .
\end{equation}
Similarly, on the closed surface $(\Sigma_2 \times\{0\})$, the topological boundary coupling can locally be written as
\begin{equation}
    -\frac{iR}{2\pi} \int_{\Sigma_2 \times \{0\}} Xdb = \frac{iR}{2\pi} \int_{\Sigma_2\times \{0\}} dX b .
\end{equation}
The full action therefore takes the following form,
\begin{equation}
    \frac{i}{2\pi} \int_{M_3} da\, b + \frac{i}{2\pi} \int_{\Sigma_2 \times \{0\}} \bigl( a+RdX \bigr)\, b -\frac{i}{2\pi} \int_{\Sigma_2 \times \{1\}} a\, b +\frac{1}{4\pi} \int_{\Sigma_2 \times \{1\}} b\, \star b. 
\end{equation}
The integral over $b$ in the interior imposes $da=0$ and the integral over its value at the topological boundary imposes $a =-R dX$. Upon collapsing the interval, this condition is imposed in the absolute theory, and the remaining action is therefore
\begin{equation}
    S_{\Sigma_2} = \frac{1}{4\pi} \int_{\Sigma_2} b\wedge \star b + \frac{iR}{2\pi} \int_{\Sigma_2} dX \wedge b.
\end{equation}
The field $b$ is auxiliary in this description. So, we can vary the action, 
\begin{equation}
    \delta_b S_{\Sigma_2} = \frac{1}{2\pi} \int_{\Sigma_2} \delta b \wedge \bigl( \star b - iRdX \bigr),
\end{equation}
and hence, 
\begin{equation}
    \star b = iR dX.
\end{equation}
Replacing this in the action $(S_{\Sigma_2})$ above, produces the action of two dimensional compact free boson at radius $R$. 
\begin{equation}
\boxed{    S_{\Sigma_2} = \frac{R^2}{4\pi} \int_{\Sigma_2} dX \wedge \star dX. }
\end{equation}
The relation $\star b = iR dX$, could have been recovered by combining the topological boundary condition $ a \big|_{\Sigma_2 \times \{0\}} = -RdX$ with the physical boundary condition $ b\big|_{\Sigma_2 \times \{1\}} =i\star a$.
This is exactly the reason why, in the literature often people perform the interval compactification by combining the boundary conditions (physical and symmetry), which produces the action of the physical theory -- in this case, the $2d$ compact free boson at radius $R$. 

\subsection{Boundary SymTFT}
To analyze the symmetries of the compact free boson theory placed on a manifold with boundary we need to use the framework of the Boundary SymTFT \cite{Bhardwaj:2024igy, Copetti:2024onh}. In other words, when we place our quantum field theory on a manifold with boundary, the physical boundary of the SymTFT is replaced by a manifold with a boundary. Consequently, we end up getting a new topological boundary and couple of corners in our SymTFT. 

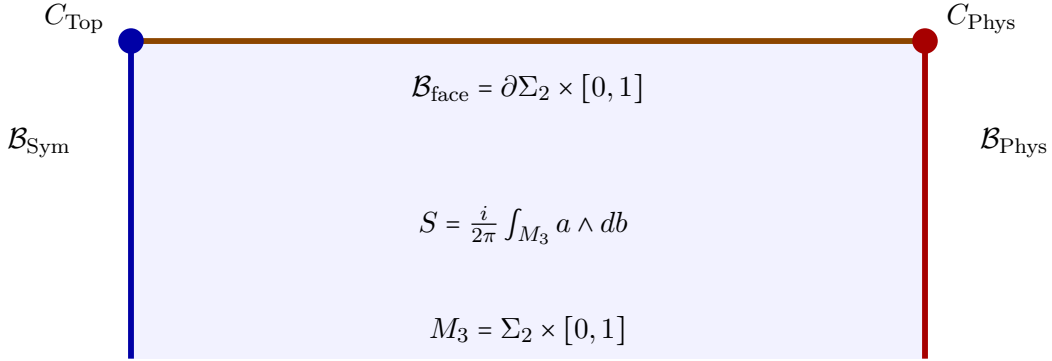
\begin{figure}[ht]
    \centering
    \begin{tikzpicture}[scale=1.5]

        \fill[blue!5] (0,0) rectangle (7,2.8);

        \draw[line width=2.2pt, blue!65!black] (0,0) -- (0,2.8);

        \draw[line width=2.2pt, red!65!black] (7,0) -- (7,2.8);

        \draw[line width=2.2pt, orange!55!black] (0,2.8) -- (7,2.8);

        \fill[blue!65!black]
            (0,2.8) circle (3.2pt);

        \fill[red!65!black]
            (7,2.8) circle (3.2pt);

        \node[anchor=north west, align=left]  at (-1.18,2.12)
            {$\mathcal{B}_{\mathrm{Sym}}$};

        \node[anchor=north east, align=right] at (8.18,2.12)
            {$\mathcal{B}_{\mathrm{Phys}}$};

        \node[align=center] at (3.5,1.18) { $S = \frac{i}{2\pi} \int_{M_3} a \wedge db$ };

        \node at (3.5,2.38) {$\mathcal{B}_{\mathrm{face}} = \partial \Sigma_2 \times [0,1]$};

        \node at (-0.5, 3.0) {$ C_{\text{Top}}$};

        \node at (7.5, 3.0) {$C_{\text{Phys}}$};

        \node[align=center]  at (3.5,0.25) {$M_3=\Sigma_2\times [0,1]$};
    \end{tikzpicture}

    \caption{When the physical spacetime $\Sigma_2$ has a boundary, there is an additional topological face $\mathcal{B}_{\mathrm{face}}$ of the SymTFT, that anchors between two corners of the physical and symmetry boundary ($C_{\text{Phys}}$ and $C_{\text{Top}}$ respectively). After reducing the sandwich aka interval compactification, the corners combine and produce the boundary of the absolute quantum field theory.}
    \label{fig:compact-boson-boundary-symtft}
\end{figure}

When $\Sigma_2$ has a boundary, $M_3=\Sigma_2\times[0,1]$ acquires a third boundary component, the topological face $\mathcal{B}_{\text{face}} = \partial\Sigma_2\times[0,1]$, which meets the symmetry boundary and the physical boundary along the corners \footnote{SymTFT's with corners have also been explored from top down String theory perspective, see \cite{Cvetic:2024dzu}.} (see Figure~\ref{fig:compact-boson-boundary-symtft}.),
\begin{equation}
    C_{\text{Top}} = \partial\Sigma_2\times\{0\} , \qquad C_{\text{Phys}} = \partial\Sigma_2 \times \{1\} \,.
\end{equation}
We orient both corners as $\partial\Sigma_2$, i.e. with the orientation induced from $\Sigma_2$, since after the interval compactification they become the boundary of the two dimensional theory. With this choice, Stokes theorem on each of the two ends of the interval reads
\begin{equation}
    \int_{\Sigma_2\times \{t\}}d\eta = \int_{\partial \Sigma_2 \times \{t\}} \eta, \qquad t=\{0,1\}.
\end{equation}
while the bulk variation acquires the additional term due the presence of the topological face.
We need to make sure the variational problem is well defined on the horizontal topological face $\mathcal{B}_{\text{face}}$, as in Figure~\ref{fig:compact-boson-boundary-symtft}. To ensure that, the boundary variation must vanish. In other words, we must impose boundary conditions to make sure 
\begin{equation}
    \delta S_{\mathrm{SymTFT}} \big|_{\mathcal{B}_{\text{face}}} = \frac{i}{2\pi}\int_{\mathcal{B}_{\text{face}}} a \delta b \rightarrow \text{needs to cancel}.
\end{equation}
This criteria can be achieved by imposing, 
\begin{equation}
    \iota^\ast a \big|_{\mathcal{B}_{\text{face}}} = 0,
\end{equation}
or we can keep $\iota^\ast  b\big|_{\mathcal{B}_{\text{face}}}$ fixed. Either choice restricts the allowed gauge transformations on the topological face.

In the presence of the corners, the variation of the symmetry boundary action produces an additional term. On the surface $\Sigma_2\times\{0\}$ with boundary $C_{\text{Top}}$, we find
\begin{equation}\label{eq: variation-sym-bdry}
    \delta S_{\text{Sym Bdry}} = \frac{iR}{2\pi} \int_{\Sigma_2 \times\{0\}} \Big[-\delta Xdb +dX  \delta b \Big] - \frac{iR}{2\pi} \int_{C_{\text{Top}}} X \delta b ,
\end{equation}
so that the symmetry boundary equations we derived earlier remains unchanged, but a corner term survives. Its fate depends on the condition we choose on the topological face, and it is what the corner edge modes (which we are going to add below) will take care of. No further corner term arises at $C_{\text{Phys}}$, as the the physical boundary action contains no derivatives, and the gauge transformations of $b$ are restricted on $\mathcal{B}_{\text{Phys}}$.

\subsubsection{Dirichlet edge mode from boundary SymTFT }

Let us first choose the condition $ \iota^\ast a \big|_{\mathcal{B}_{\text{face}}} = 0$. Therefore, we have three conditions on three separate boundaries to the bulk theory. 
\begin{equation}
    \begin{aligned}
        &1. ~ a = - R dX ,\quad  \text{on} ~ \mathcal{B}_{\text{Sym}}, \\
        &2. ~ a =i \star b, \quad \text{on} ~ \mathcal{B}_{\text{Phys}}, \\
        &3. ~ a = 0 , \quad \text{on} ~ \mathcal{B}_{\text{face}}.
    \end{aligned}
\end{equation}
At the corner $C_{\text{Top}}$, both 1. and 3. should be compatible. Hence, combining them we get, 
\begin{equation}
    dX \big|_{C_{\text{Top}}} =0.
\end{equation}
We can realize this condition by introducing an edge mode localized on the corner $C_{\text{Top}}$. We add, 
\begin{equation}\label{eq: corner-edge-dirichlet}
    S_{C_{\text{Top}}} = \frac{i}{2\pi} \int_{C_{\text{Top}}}  \Psi  dX + \frac{iR}{2\pi} \int_{C_{\text{Top}}}X\,b.
\end{equation}
where $\Psi$ is a $U(1)$-valued scalar localized on the corner. The second term is a local counterterm. The counterterm amounts to writing the symmetry boundary action in the form $\frac{iR}{2\pi} \int_{\Sigma_2\times\{0\}} dX\wedge b$ (instead of $X\,db$ in the integrand). Moreover, its variation, $\frac{iR}{2\pi} \int_{C_{\text{Top}}}(\delta X\,b+X\,\delta b)$, trades the $X\,\delta b$ term in \eqref{eq: variation-sym-bdry} for a $\delta X\,b$ term.

The variation of the full action around $C_{\text{Top}}$ is 
\begin{equation}
    -\frac{iR}{2\pi} \int_{C_{\text{Top}}} X\delta b + \frac{iR}{2\pi}\int_{C_{\text{Top}}} \big(\delta Xb+X\delta b \big) + \frac{i}{2\pi}\int_{C_{\text{Top}}} \big(\delta\Psi dX - d\Psi \delta X\big).
\end{equation}
This expression can be massaged into the following, 
\begin{equation}
    \frac{i}{2\pi}\int_{C_{\text{Top}}}\Big[ \delta \Psi dX + \delta X \big(R b - d\Psi \big) \Big].
\end{equation}
The corner equations of motion are therefore
\begin{equation}\label{eq:corner-eom-D}
\Psi:\ dX \big|_{C_{\text{Top}}} = 0 , \qquad X:\ R b\big|_{C_{\text{Top}}} = d\Psi ,
\end{equation}
the first of which is the desired compatibility condition.

One might be worried about the gauge invariance now, but this corner action is gauge invariant on its own due to the following. We have chosen $\iota^\ast a =0$ on the topological face $\mathcal{B}_{\text{face}}$, which restricts the allowed gauge transformations on the topological face. We require, 
\begin{equation}
    d\lambda_a \big|_{\mathcal{B}_{\text{face}}} = 0.
\end{equation}
Under $X \rightarrow X - R^{-1}\lambda_a$, at the corner $C_{\text{Top}}$, 
\begin{equation}
    dX \big|_{C_{\text{Top}}} \rightarrow dX \big|_{C_{\text{Top}}} - R^{-1} d\lambda_a = dX \big|_{C_{\text{Top}}}. 
\end{equation}
The counterterm is invariant under $b\to b+d\lambda_b$ provided $\Psi$ transforms as
\begin{equation}
\Psi \rightarrow \Psi + R\lambda_b .
\end{equation}
Consequently we do not have to worry about the gauge invariance of the corner action. 

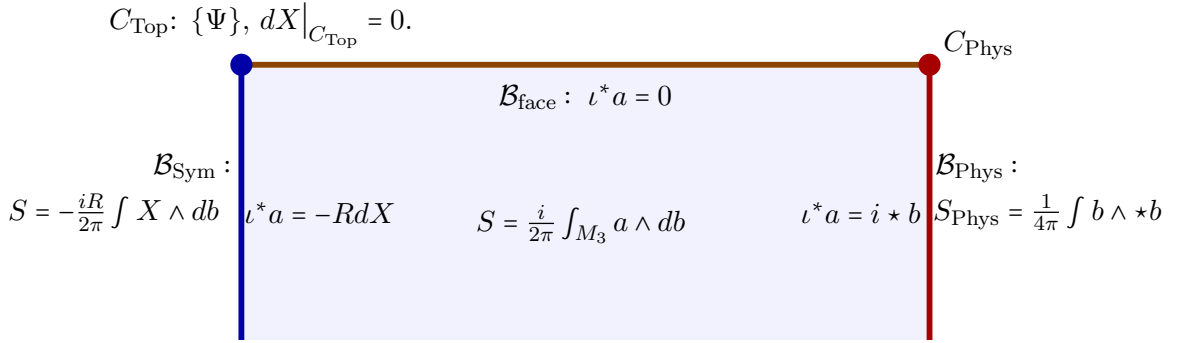
\begin{figure}[ht]
    \centering
    \begin{tikzpicture}[scale= 1.3]

        \fill[blue!5] (0,0) rectangle (7,2.8);

        \draw[line width=2.2pt, blue!65!black] (0,0) -- (0,2.8);

        \draw[line width=2.2pt, red!65!black] (7,0) -- (7,2.8);

        \draw[line width=2.2pt, orange!55!black] (0,2.8) -- (7,2.8);

        \fill[blue!65!black]
            (0,2.8) circle (3.2pt);

        \fill[red!65!black]
            (7,2.8) circle (3.2pt);

        \node[anchor=north west, align=left]  at (-1.0,2.0)
            {$\mathcal{B}_{\mathrm{Sym}}:$};

        \node at (-1.3,1.3) {$ S = -\frac{iR}{2\pi} \int X \wedge db$};

        \node at (0.8, 1.3) {$ \iota^\ast a= - R dX$};

        \node[anchor=north east, align=right] at (8.0,2.0)
            {$\mathcal{B}_{\mathrm{Phys}}: $};
            
        \node at (8.2, 1.3) {$S_{\text{Phys}}= \frac{1}{4\pi} \int b \wedge \star b$};

        \node at (6.3,1.3) {$\iota^\ast a = i \star b$};

        \node[align=center] at (3.5,1.18) { $S = \frac{i}{2\pi} \int_{M_3} a \wedge db$ };

        \node at (3.5,2.48) {$\mathcal{B}_{\mathrm{face}}: ~ \iota^\ast a =0$};

        \node at (0.2, 3.2) {$ C_{\text{Top}}$:  $ \{\Psi\} $, $dX\big|_{C_{\text{Top}}} =0$.};

        \node at (7.5, 3.0) {$C_{\text{Phys}}$};
        
    \end{tikzpicture}

    \caption{We have shown all the boundary actions and the boundary conditions. Compatibility at $C_{\mathrm{Top}}$ is implemented by the corner edge mode $\Psi$. After interval compactification, the corner data reproduce the Dirichlet edge mode theory on $\partial\Sigma_2$.}
    \label{fig:compact-boson-boundary-symtft-2}
\end{figure}

We are going to perform the interval compactification now; the situation is summarized in Figure~\ref{fig:compact-boson-boundary-symtft-2}.
 The boundary conditions on the two ends are
\begin{equation}
    a \big|_{\Sigma_2 \times \{0\} }= -R dX, \qquad a\big|_{\Sigma_2 \times \{1\} }  = i\star b,
\end{equation}
and we combine them as before to produce the action of the compact free boson. 
\begin{equation}
\boxed{  \frac{iR}{2\pi}\int_{\Sigma_2 \times \{0\}}dX\wedge b+ \frac{1}{4\pi}\int_{\Sigma_2\times \{1\} } b \wedge \star b \xrightarrow{a= -R dX, \, a = i \star b}  \frac{R^2}{4\pi} \int_{\Sigma_2} dX \wedge \star dX. }
\end{equation}
The condition on the corner $C_{\text{Top}}$ i.e. $dX \big|_{C_\text{Top}}=0$,
\begin{equation}
\boxed{    dX \big|_{C_\text{Top}}=0 \xrightarrow{\text{interval compactification}} dX \big|_{\partial \Sigma_2} = 0, }
\end{equation}
reproduces the boundary equations of motion we derived earlier in section 2. The other corner equation $Rb =  d\Psi$,
\begin{equation}
\boxed{    Rb = d\Psi \xrightarrow{a= -R dX, \, a = i\star b } R^2 \star dX = id\Psi. }
\end{equation}
In other words, these are same boundary equations of motion we derived in section \ref{sec: free boson boundary} by adding the following edge mode action,
\begin{equation}
    S^{\text{Dirichlet}}_{\text{edge mode}} = \frac{i}{2\pi} \int_{\partial \Sigma_2} \Psi  dX.
\end{equation}
Therefore, one can say that the corners combine to produce the edge mode action we used to impose Dirichlet boundary conditions on the boundary of our absolute theory.

\subsubsection{ Neumann edge mode from boundary SymTFT}

Our next target is to reproduce the boundary equation of motions in equation \eqref{eq: Neumann-edge-mode-equations} from the boundary SymTFT formalism. The essential difference from the Dirichlet construction is the choice of polarization on the horizontal topological face. Recall that the restriction of the bulk variation to $\mathcal B_{\mathrm{face}}$ is
\begin{equation}
\delta S_{\mathrm{SymTFT}}\big|_{\mathcal B_{\mathrm{face}}}
=
\frac{i}{2\pi} \int_{\mathcal B_{\mathrm{face}}} a\,\delta b. 
\end{equation}
In the Dirichlet construction, we used $\iota^\ast a \vert_{\mathcal{B}_{\text{face}}} =0$ to ensure the boundary variation vanishes. However, the variation also goes away if we keep $b$ fixed on the face. Without losing much generality, we are going to choose $\iota^\ast b=0$. Eventually, we are going to reproduce show that this choice is compatible with the corner conditions. Consequently, the gauge transformations of the field also gets restricted.   

The boundary conditions on the symmetry boundary and the physical boundary remain unchanged. 
\begin{equation}
      a \big|_{\Sigma_2 \times \{ 0\}} = -R dX, \qquad db\big|_{\Sigma_2 \times \{ 0\}} =0, \qquad b \big|_{\Sigma_2 \times \{ 1\}} = i\star a.
\end{equation}
Since the boundary condition on the topological face has changed, the boundary conditions on the symmetry boundary and on the topological face do not produce $dX \big|_{C_{\text{Top}}} =0$, on the corner. In other words, the symmetry boundary condition and the boundary condition on the face are compatible with each other even without requiring $dX \big|_{C_{\text{Top}}} =0$. This is precisely what we expect for a Neumann boundary conditions.

In order to reproduce the Neumann edge mode theory, we introduce as before, edge modes on the corner,
\begin{equation}
    S_{C_{\text{Top}}} = \frac{i}{2\pi} \int_{C_{\text{Top}}} \big( \Psi_2 \, dX + \Psi_1 \, d \Psi_2 \big) + \frac{iR}{2\pi} \int_{C_{\text{Top}}} X\,b,
\end{equation}
with the same counterterm as in \eqref{eq: corner-edge-dirichlet}.
The variation of the full corner action is now, 
\begin{equation}
   \delta S\big|_{C_{\text{Top}}} = \frac{i}{2\pi} \int_{C_{\text{Top}}} \Big[ \delta\Psi_1 \,d\Psi_2 + \delta \Psi_2\, \big( dX - d\Psi_1 \big) + \delta X\, \big(R b - d\Psi_2 \big) \Big] .
\end{equation}
Note that, since $\delta b =0$ now, we do not pick up any additional contribution to the variation above from the symmetry boundary action \eqref{eq: sym-boundary-edge}. The corner equations of motion are, 
\begin{equation}
   \Psi_1:\ d\Psi_2=0 \qquad \Psi_2:\ dX\big|_{C_{\text{Top}}} = d\Psi_1,\qquad X:\ Rb|_{C_{\text{Top}}} = d\Psi_2.
\end{equation}
Combining the first and the last equation we get $b\big|_{C_{\text{Top}}} = 0$, which is exactly what the condition we imposed on the topological face earlier. Therefore, the corner equations are compatible with the face.

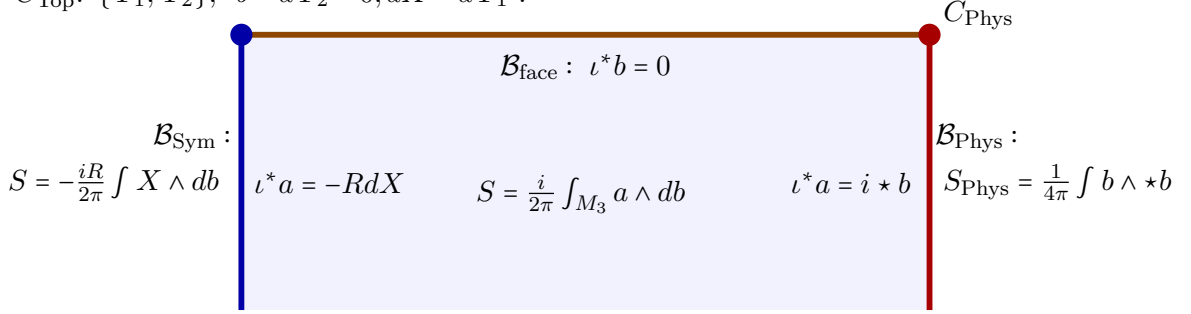
\begin{figure}[ht]
    \centering
    \begin{tikzpicture}[scale= 1.3]

        \fill[blue!5] (0,0) rectangle (7,2.8);

        \draw[line width=2.2pt, blue!65!black] (0,0) -- (0,2.8);

        \draw[line width=2.2pt, red!65!black] (7,0) -- (7,2.8);

        \draw[line width=2.2pt, orange!55!black] (0,2.8) -- (7,2.8);

        \fill[blue!65!black]
            (0,2.8) circle (3.2pt);

        \fill[red!65!black]
            (7,2.8) circle (3.2pt);

        \node[anchor=north west, align=left]  at (-1.0,2.0)
            {$\mathcal{B}_{\mathrm{Sym}}:$};

        \node at (-1.3,1.3) {$ S = -\frac{iR}{2\pi} \int X \wedge db$};

        \node at (0.9, 1.3) {$ \iota^\ast a= - R dX$};

        \node[anchor=north east, align=right] at (8.0,2.0)
            {$\mathcal{B}_{\mathrm{Phys}}: $};
            
        \node at (8.3, 1.3) {$S_{\text{Phys}}= \frac{1}{4\pi} \int b \wedge \star b$};

        \node at (6.2,1.3) {$\iota^\ast a = i \star b$};

        \node[align=center] at (3.5,1.18) { $S = \frac{i}{2\pi} \int_{M_3} a \wedge db$ };

        \node at (3.5,2.48) {$\mathcal{B}_{\mathrm{face}}: ~ \iota^\ast b = 0$};

        \node at (0.3, 3.2) {$ C_{\text{Top}}$:  $\{\Psi_1, \Psi_2\} $,~ $b =d\Psi_2=0, dX =d\Psi_1$ .};

        \node at (7.5, 3.0) {$C_{\text{Phys}}$};
        
    \end{tikzpicture}

    \caption{We have shown all the boundary actions and the boundary conditions. Compatibility at $C_{\mathrm{Top}}$ is implemented by the corner edge mode $\Psi_1$ and $\Psi_2$. After interval compactification, the corner data reproduce the Neumann edge mode theory on $\partial\Sigma_2$.}
    \label{fig:compact-boson-boundary-symtft-3}
\end{figure}

Reducing the sandwich at this stage, produces the action of the compact free boson on $\Sigma_2$. Beyond that, on $\partial \Sigma_2$ we have,  
\begin{equation}
\boxed{    dX \big|_{C_\text{Top}}= d\Psi_1 ,\, d\Psi_2 =0 \xrightarrow{\text{interval compactification}} dX \big|_{\partial \Sigma_2} = d\Psi_1, \, d\Psi_2 =0. }
\end{equation}
Moreover, the other corner equation $Rb = d\Psi_2= 0$,
\begin{equation}
\boxed{    Rb\big|_{C_{\text{Top}}} =d\Psi_2 = 0 \xrightarrow{a= -R dX, \quad a =i \star b } R^2 \star dX \big|_{\partial \Sigma_2} = 0. }
\end{equation}
These are the same boundary equations of motion we obtained for our Neumann edge mode action in Section~\ref{sec: free boson boundary} (see Figure~\ref{fig:compact-boson-boundary-symtft-3} for an illustration.). 

\subsection{Codimension one defects of the free boson SymTFT}

The compact free boson at radius $R$ is T-dual to the compact free boson at radius $1/R$ and it is a classic fact that T-duality exchanges the Dirichlet and Neumann boundary conditions. In the language of the previous subsection, T-duality should be visible as the action of codimension-one defect of the SymTFT on the boundary data of the boundary SymTFT: the symmetry boundary, the topological face, and the corners. The relevant defect was constructed in \cite{Argurio:2024ewp} (see also, \cite{Niro:2022ctq, Thorngren:2021yso} for the T-duality defects of the compact boson from a purely two dimensional point of view). Since the construction of \cite{Argurio:2024ewp} is formulated for the closed sandwich, i.e. for $\Sigma_2$ without boundary, we first review the construction, and then we use it in the presence of boundaries and corners. 

\subsubsection{The T-duality defect}

Let us go back to the original SymTFT action $S_{\text{SymTFT}} = \frac{i}{2\pi}\int_{M_3} a db$ and for a moment forget the topological face. Since $a$ and $b$ are $\mathbb{R}$-valued, they may be rescaled by an arbitrary real number without violating any quantization condition. The bulk action is invariant under this one-parameter family of transformations\footnote{There are other such symmetries of the bulk TFT, for example scaling. See \cite{Gaiotto:2020iye} for further details.}
\begin{equation}\label{eq: bulk-duality-map}
 \boxed{   a\rightarrow sb , \qquad b \rightarrow s^{-1} a, \qquad s\in \mathbb{R}_{\neq 0}. }
 \end{equation}
Indeed, the parameter $s$ drops out immediately,
\begin{equation}\label{eq: bulk-duality-invariance}
    \frac{i}{2\pi}\int_{M_3} (sb)  d(s^{-1}a) = \frac{i}{2\pi}\int_{M_3} b da = \frac{i}{2\pi} \int_{M_3} a  db + \frac{i}{2\pi}\int_{\partial M_3} ab.
\end{equation}
So \eqref{eq: bulk-duality-map} is a symmetry of the bulk theory on a closed manifold, and on a manifold with boundary it is a symmetry up to a boundary term, which will be important later. Such symmetries of the bulk theory will be generated by codimension one defects. Within this family of defects generated by \eqref{eq: bulk-duality-map}, special elements are, 
\begin{equation}
    s= 1, ~ (a,b) \mapsto (b,a) , \qquad s= -1, ~ (a,b) \mapsto (-b,-a). 
\end{equation}
We will see that the defect with $s=1$ corresponds to T-duality. In \cite{Argurio:2024ewp} (See \cite{Robbins:2025urk} for a similar construction), this defect was constructed as condensation defect using higher gauging techniques \cite{Roumpedakis:2022aik,Gaiotto:2019xmp,Lin:2022xod}. However, we are going to using the defect action method (gluing method) for our purposes \cite{Choi:2021kmx, Choi:2022zal, Choi:2022jqy}. 

Let $D \subset M_3$ be a surface which separates $M_3$ into two pieces, $M_3 = M_L \cup_D M_R$, and let us denote the fields on the two sides by $(a_L,b_L)$ and $(a_R,b_R)$. We orient $D$ as (part of) the boundary of $M_L$, 
\begin{equation}
\int_{M_L} d\omega = \int_D \omega + \ldots,  \qquad \int_{M_R} d\omega = -\int_D \omega + \ldots.
\end{equation}
When placed along $D = \Sigma_2 \times \{t_0\}$, such that $M_L = \Sigma_2 \times [0,t_0]$, and $M_R = \Sigma_2 \times [t_0,1]$, the defect $\mathcal{T}_S[D]$ can be defined by the gluing conditions of the fields on each side, 
\begin{equation}\label{eq: gluing conditions}
\boxed{  \mathcal{T}_s[D]: \qquad  a_R = s  b_L , \qquad b_R = s^{-1} a_L.  }
\end{equation}
The gauge parameters are glued in the same way, 
\begin{equation}
    \lambda_a^{R} = s \lambda_b^{L}, \qquad \lambda_b^{R} = s^{-1}\lambda_a^{L}.
\end{equation}
\begin{figure}[ht]
    \centering
    \begin{tikzpicture}[scale=1.5]
        \fill[blue!5] (0,0) rectangle (7,2.8);
        \draw[line width=2.2pt, red!65!black] (0,0) -- (0,2.8);
        \draw[line width=2.2pt, blue!65!black] (7,0) -- (7,2.8);
        \draw[line width=1.8pt, green!45!black, dashed] (3.5,0) -- (3.5,2.8);
        \node[green!45!black] at (3.5,3.05) {$\mathcal{T}_s$};
        \draw[line width=1pt, purple!70!black] (5.6,1.4) -- (3.5,1.4);
        \draw[line width=1pt, purple!70!black, densely dotted] (3.5,1.4) -- (1.4,1.4);
        \fill[purple!70!black] (3.5,1.4) circle (2pt);
        \node[purple!70!black, anchor=south] at (4.55,1.45) {$U_m$};
        \node[purple!70!black, anchor=south] at (2.45,1.45) {$V_{sm}$};
        \node at (1.75,0.35) {$(a_L,b_L)$};
        \node at (5.25,0.35) {$(a_R,b_R)$};
        \node[green!45!black, anchor=west] at (3.65,2.55) {\footnotesize $a_R = s\, b_L$};
        \node[green!45!black, anchor=west] at (3.65,2.25) {\footnotesize $b_R = s^{-1} a_L$};
        \node[anchor=south] at (0,2.9) {$\mathcal{B}_{\text{Sym}}:~ a = -R\, dX$};
        \node[anchor=south] at (7,2.9) {$\mathcal{B}_{\text{Phys}}:~ a = i \star b$};
        \node at (3.5,-0.3) {$M_3 = \Sigma_2 \times [0,1]$};
    \end{tikzpicture}
    \caption{The duality defect $\mathcal{T}_s$ placed vertically in the sandwich, at $\Sigma_2 \times \{t_0\}$. The fields on the two sides satisfy the gluing conditions on the defect. A line $U_m$ crossing the defect continues as $V_{sm}$ on the other side. Pushing the defect onto the physical boundary leaves it invariant for $s=\pm 1$, while pushing it onto the symmetry boundary maps $\mathcal{B}^{(R)}_{\text{Sym}}$ to $\mathcal{B}^{(s/R)}_{\text{Sym}}$.}
    \label{fig: T-duality defect}
\end{figure}
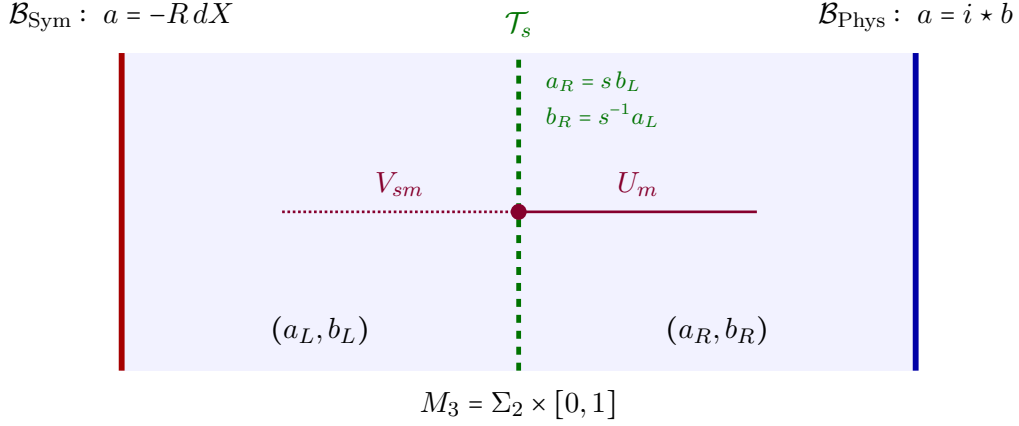
We determine the action living on $D$ by demanding a well-posed variational problem. Varying the bulk action on the two pieces, we find the following terms on $D$,
\begin{equation}
    \delta \Big( \frac{i}{2\pi} \int_{M_L} a_L  db_L + \frac{i}{2\pi}\int_{M_R} a_R db_R \Big) \Big|_{D} = -\frac{i}{2\pi}\int_D a_L \delta b_L + \frac{i}{2\pi} \int_D a_R \delta b_R,
\end{equation}
using the gluing conditions, this expression can be massaged into the following 
\begin{equation}
    -\frac{i}{2\pi}\int_D \Big( \delta a_L  b_L + a_L \delta b_L \Big) = -\frac{i}{2\pi} \delta \int_D a_L b_L.
\end{equation}
Therefore the defect must carry the action
\begin{equation}\label{eq: defect-action}
\boxed{  S_{\mathcal{T}_s} = \frac{i}{2\pi} \int_D a_L  b_L = - \frac{i}{2\pi}\int_D a_R  b_R . }
\end{equation}
It is clear that the defect action is gauge invariant on a closed $D$. When $D$ has a boundary, as will be the case in the boundary SymTFT, the total derivative terms produce contributions on the boundary of the defect worldvolume, we will deal with them later.

Moreover, the gluing conditions \eqref{eq: gluing conditions} also determine what happens to a line operator when it `crosses' the defect (see Figure~\ref{fig: T-duality defect}). When, $U_m[\gamma]$ crosses $D$ transversally from $M_L$ into $M_R$, it transforms into $V_{sm}[\gamma]$. In other words, the operator $\exp(im\int a_L)$ continues across $D$ as $\exp(ims \int b_R)$. So, we can write
\begin{equation}\label{eq: species change}
\boxed{  U_m \big|_{M_L} \longleftrightarrow V_{sm}\big|_{M_R}, \qquad V_n\big|_{M_L} \longleftrightarrow U_{n/s}\big|_{M_R}.}
\end{equation}

\subsubsection{Action on the boundaries}

Since $\mathcal{T}_s$ is topological, we can push it onto either boundary, and the question we would like to ask is what happens to the boundary conditions after we fuse the defect on top of the boundaries. 

\paragraph{Physical boundary.} To begin with, we are going to investigate how the physical boundary changes under the action of the defect. Recall that the physical boundary imposes $a = i\star b$. Once we push $\mathcal{T}_s$ onto the physical boundary, i.e.~send $t_0\rightarrow 1$, we find 
\begin{equation}\label{eq: physical-boundary-change}
    a_R = i \star b_R \quad \longrightarrow \quad sb_L = i \star  s^{-1} a_L \quad \longleftrightarrow \quad b_L = s^{-2} i \star a_L.
\end{equation}
Therefore the physical boundary condition is invariant precisely for $s^2 = 1$, as pointed out in \cite{Argurio:2024ewp}. 

\paragraph{Symmetry boundary.} Now push $\mathcal{T}_s$ onto the symmetry boundary at $t_0 = 0$. All the fields are now the $(a_R,b_R)$ ones, hence we are going to drop the subscript unless needed. The symmetry boundary action \eqref{eq: sym-boundary-edge}, written in terms of $b_L$, 
\begin{equation}
    S_{\text{Sym Bdry}} = - \frac{iR}{2\pi} \int_{\Sigma_2 \times \{ 0 \}} X\, db_L,
\end{equation}
and the defect action~\eqref{eq: defect-action}, now sitting on $\Sigma_2 \times \{0\}$, combine to give us new symmetry boundary action.
\begin{equation}\label{eq: dual symmetry boundary action}
\begin{aligned}
    S_{\text{Sym Bdry}}' &= -\frac{iR}{2\pi} \int_{\Sigma_2\times\{0\}} X d\big( s^{-1} a \big) + \frac{i}{2\pi} \int_{ \Sigma_2 \times \{0\} } (s b)  (s^{-1}a) \\
    &= -\frac{iR}{2\pi s} \int_{\Sigma_2\times\{0\}} X da + \frac{i}{2\pi} \int_{\Sigma_2 \times \{0\}} b  a ,
\end{aligned}
\end{equation}
where the edge mode $X$ is the same $2\pi$-periodic scalar as before, and its gauge transformation gets modified, from $X \rightarrow X - R^{-1}\lambda_a^{L}$ to $X \rightarrow X - (s/R) \lambda^R_b$. Once again, we can identify the new boundary condition at the symmetry boundary. By doing a classical variation we are left with
\begin{equation}\label{eq: dual-symmetry boundary-variation}
    \delta S\big|_{\Sigma_2 \times \{0\}} = -\frac{i}{2\pi}\int_{\Sigma_2\times\{0\}} \Big[ \delta a \Big( b + \frac{R}{s}\, dX \Big) + \frac{R}{s} \delta X da \Big] ,
\end{equation}
where we have added the contribution from the bulk $ \frac{i}{2\pi}\int_{\Sigma_2\times\{0\}} a\,  \delta b$.
So the new symmetry boundary conditions are
\begin{equation}\label{eq: dual-symmetry boundary-conditions}
    \boxed{ b \big|_{ \Sigma_2 \times\{0\}} = - \frac{R}{s} dX, \qquad da \big|_{ \Sigma_2 \times \{0\} } = 0 . }
\end{equation}
Comparing with the original symmetry boundary, we see that the roles of $a$ and $b$ have been exchanged, as expected, and that $R$ has been replaced by $R/s$.

Beyond the local equations of motion, there is some global information on the symmetry boundary. In the original boundary action, the sum over the winding sectors of $X$ in $-\frac{iR}{2\pi}\int X db = \frac{iR}{2\pi} \int dX b$ forces the holonomies of $b$ to be quantized, $\oint b \in \frac{2\pi}{R}\mathbb{Z}$. In the same way, the sum over the windings of $X$ in the new symmetry boundary action, enforces, 
\begin{equation}
    \oint a \in \frac{2\pi s}{R} \mathbb{Z}. 
\end{equation}
In addition to this we have, $da=0$. Both of these conditions allow us to introduce a new $2\pi$-periodic scalar $X'$ -- the dual scalar -- with the following constraint,
\begin{equation}\label{eq: dual edge mode}
    a \big|_{ \Sigma_2 \times\{0\} } = -\frac{s}{R} dX', \qquad X' \sim X' + 2\pi, \qquad X' \rightarrow X' - \frac{R}{s} \lambda_a ,
\end{equation}
imposing this condition on the new symmetry boundary condition is equivalent to integrating out $X$. That leaves us with only the second term in \eqref{eq: dual symmetry boundary action}. 
\begin{equation}\label{eq: dual edge symmetry}
    S'_{\text{Sym Bdry}} = \frac{i}{2\pi} \int_{\Sigma_2 \times \{ 0\}} b \,(\frac{-s}{R} dX') = -\frac{is}{2\pi R} \int_{\Sigma_2 \times \{ 0\}} b\,  dX' = \frac{i(s/R)}{2\pi} \int_{\Sigma_2 \times \{0\}} X'\, db.
\end{equation}
This is precisely the symmetry boundary action with $R$ replaced by $s/R$ and $X$ replaced by $X'$.
The boundary condition \eqref{eq: dual-symmetry boundary-conditions}, can be rephrased as $a = -R'dX'$ at $R' = s/R$, and the gauge transformation is $X'\rightarrow X' - R'^{-1}\lambda_a$. Denoting by $\mathcal{B}^{(R)}_{\text{Sym}}$ the symmetry boundary with edge mode coupling \eqref{eq: sym-boundary-edge} at radius $R$, we have shown
\begin{equation}\label{eq: action-of-T_s-on-symmetry boundary}
\boxed{ \mathcal{T}_s \otimes  \mathcal{B}^{(R)}_{\text{Sym}} = \mathcal{B}^{(s/R)}_{\text{Sym}} , \qquad \text{in particular} \qquad \mathcal{T}_1 \otimes \mathcal{B}^{(R)}_{\text{Sym}} = \mathcal{B}^{(1/R)}_{\text{Sym}}. }
\end{equation}
Now, if we focus only on $\mathcal{T}_1$, closing the sandwich with the defect in the middle (vertically), produces the compact free boson theory at radius $1/R$. This is the SymTFT realization of T-duality. Next, we are going to discuss this defect in the boundary SymTFT setting. 

\subsection{T-duality exchanging Dirichlet and Neumann edge theories}
\begin{figure}[ht]
    \centering
    \begin{tikzpicture}[scale=1.5]
        \fill[blue!5] (0,0) rectangle (7,2.8);
        \draw[line width=2.2pt, red!65!black] (0,0) -- (0,2.8);
        \draw[line width=2.2pt, blue!65!black] (7,0) -- (7,2.8);
        \draw[line width=2.2pt, orange!55!black] (0,2.8) -- (7,2.8);
        \draw[line width=1.8pt, green!45!black, dashed] (3,0) -- (3,2.8);
        \fill[green!45!black] (3,2.8) circle (2.6pt);
        \fill[red!65!black] (0,2.8) circle (3.2pt);
        \fill[blue!65!black] (7,2.8) circle (3.2pt);
        \node[green!45!black] at (3,-0.3) {$\mathcal{T}_s$ at $\Sigma_2\times\{t_0\}$};
        \node[green!45!black, anchor=south] at (3,2.95) {$\partial D$};
        \node[orange!55!black, anchor=south] at (1.5,2.95) {$\mathcal{B}^L_{\text{face}}:~\iota^\ast a_L = 0$};
        \node[orange!55!black, anchor=south] at (5.1,2.95) {$\mathcal{B}^R_{\text{face}}:~\iota^\ast b_R = 0$};
        \node[anchor=south east] at (-0.1,3.05) {$C_{\text{Top}}:\{\Psi\}$};
        \node[anchor=south west] at (7.1,3.05) {$C_{\text{Phys}}$};
        \node[anchor=east] at (-0.15,1.4) {$\mathcal{B}^{(R)}_{\text{Sym}}$};
        \node[anchor=west] at (7.15,1.4) {$\mathcal{B}_{\text{Phys}}$};
        \node at (1.5,1.4) {$(a_L,b_L)$};
        \node at (5.0,1.4) {$(a_R,b_R)$};
        \node[green!45!black, anchor=west] at (3.1,2.35) {\footnotesize $a_R = sb_L$};
        \node[green!45!black, anchor=west] at (3.1,2.05) {\footnotesize $b_R = s^{-1}a_L$};
        \node at (-0.3,0.3) {(a)};
        \begin{scope}[yshift=-4.6cm]
        \fill[blue!5] (0,0) rectangle (7,2.8);
        \draw[line width=2.2pt, red!65!black] (0,0) -- (0,2.8);
        \draw[line width=2.2pt, blue!65!black] (7,0) -- (7,2.8);
        \draw[line width=2.2pt, orange!55!black] (0,2.8) -- (7,2.8);
        \fill[red!65!black] (0,2.8) circle (3.2pt);
        \fill[blue!65!black] (7,2.8) circle (3.2pt);
        \node[orange!55!black, anchor=south east] at (4.5,2.95) {$\mathcal{B}_{\text{face}}:~\iota^\ast b = 0$};
        \node[anchor=south west] at (-1.55,3.05) {$C_{\text{Top}}:\{\Psi_1 = -\Psi,\ \Psi_2 = -X|_{C_{\text{Top}}}\}$};
        \node[anchor=south west] at (7.1,3.05) {$C_{\text{Phys}}$};
        \node[anchor=east] at (-0.15,1.4) {$\mathcal{B}^{(s/R)}_{\text{Sym}}$};
        \node[anchor=west] at (7.15,1.4) {$\mathcal{B}_{\text{Phys}}$};
        \node at (3.5,1.6) {$a = -\tfrac{s}{R}\, dX'$ on $\Sigma_2\times\{0\}$};
        \node at (3.5,1.1) {$\mathcal{T}_s$ pushed onto $\mathcal{B}_{\text{Sym}}$, edge mode dualized};
        \node at (-0.3,0.3) {(b)};
        \end{scope}
    \end{tikzpicture}
    \caption{(a) The vertical defect in the boundary SymTFT, ending on the face along $\partial D = \partial\Sigma_2\times\{t_0\}$. The face is of Dirichlet type on the left side and of Neumann type on the right side of the defect. (b) After pushing the defect onto the symmetry boundary and dualizing the edge mode, the configuration is the Neumann one at radius $s/R$, and the Dirichlet corner mode $\Psi$ together with the boundary value of $X$ have become the two Neumann corner modes.}
    \label{fig: vertical defect}
\end{figure}
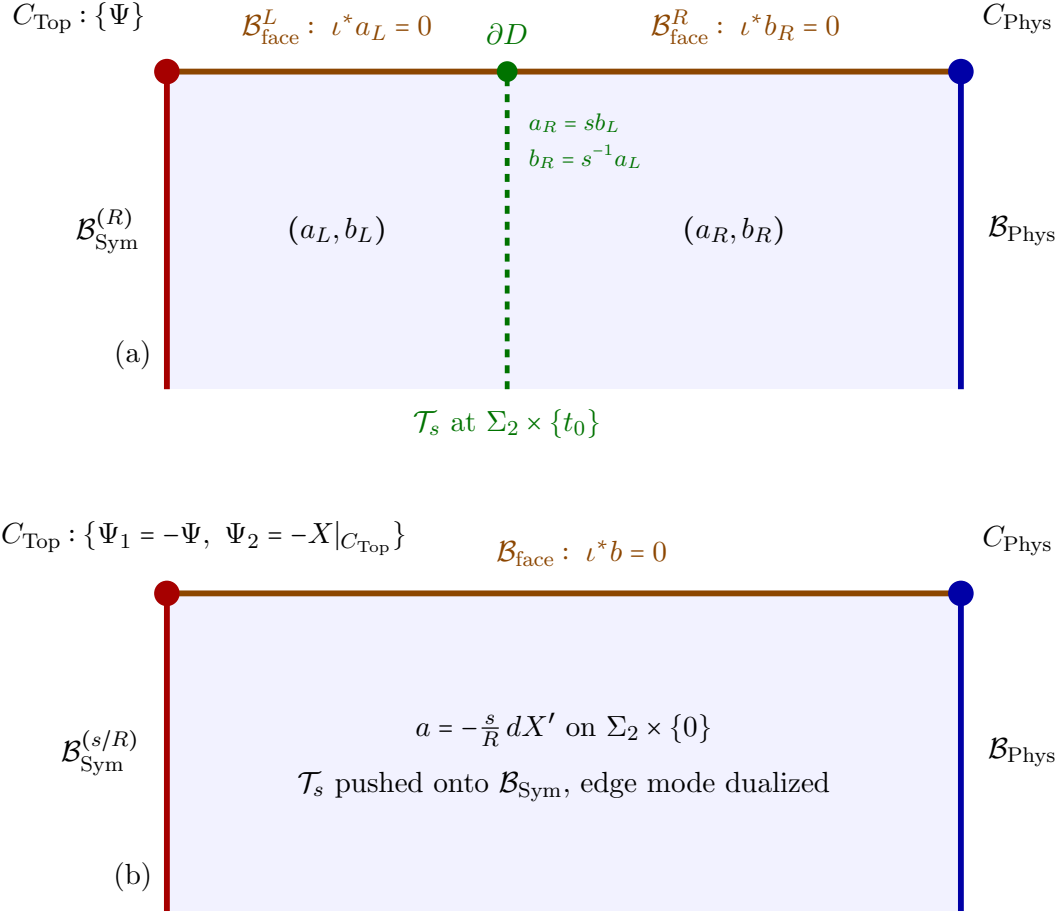

In the boundary SymTFT environment, when we place the defect $\mathcal{T}_s[D]$ vertically in the middle, where $D = \Sigma_2 \times \{t_0\}$, the first change we see is that the defect worldvolume now has a boundary. 
\begin{equation}
    \partial D = \partial \Sigma_2 \times \{t_0\},
\end{equation}
which lies on the topological face, $\mathcal{B}_{\text{face}}$. Therefore, the face is now cut into two pieces (see Figure~\ref{fig: vertical defect} for an illustration.), 
\begin{equation}
    \mathcal{B}^L_{\text{face}} = \partial \Sigma_2 \times [0,t_0], \qquad \mathcal{B}^R_{\text{face}} = \partial \Sigma_2 \times [t_0,1]. 
\end{equation}
Let us first remind ourselves about the two configurations (Dirichlet and Neumann) we found in the previous subsection. Both have the same symmetry boundary $\mathcal{B}^{(R)}_{\text{Sym}}$, with action \eqref{eq: sym-boundary-edge}, and the same physical boundary; they differ in the condition on the topological face and in the corner theory:
\begin{equation}\label{eq: D and N configurations}
\begin{aligned}
    \mathsf{D}_R:& \qquad \iota^\ast a\big|_{\mathcal{B}_{\text{face}}} = 0, \qquad S_{C_{\text{Top}}} = \frac{i}{2\pi}\int_{C_{\text{Top}}} \Psi\, dX + \frac{iR}{2\pi}\int_{C_{\text{Top}}} X \,b, \\
    \mathsf{N}_R:& \qquad \iota^\ast b\big|_{\mathcal{B}_{\text{face}}} = 0, \qquad S_{C_{\text{Top}}} = \frac{i}{2\pi}\int_{C_{\text{Top}}} \big( \Psi_2 dX + \Psi_1 d\Psi_2 \big) + \frac{iR}{2\pi}\int_{C_{\text{Top}}} X\, b.
\end{aligned}
\end{equation}
As we mentioned, the defect $\mathcal{T}_s[D]$ cuts the face in two pieces. We start from $\mathsf{D}_R$ on the left, so on $\mathcal{B}^L_{\text{face}}$ we keep $\iota^\ast a_L = 0$. On $\mathcal{B}^R_{\text{face}}$ we impose the condition obtained from it by the gluing map \eqref{eq: gluing conditions}, $a_L = s b_R$,
\begin{equation}\label{eq: face conditions with defect}
    \iota^\ast a_L \big|_{\mathcal{B}^L_{\text{face}}} = 0, \qquad \iota^\ast b_R \big|_{\mathcal{B}^R_{\text{face}}} = 0.
\end{equation}
So the face is of Dirichlet type on the left side of the defect and of Neumann type on the right. 

We would like to emphasize that no additional degrees of freedom and no additional conditions are needed on the line $\partial D$. Each piece of the face kills its own boundary variation, which ensures variational problem is well defined. Moreover, the defect action \eqref{eq: defect-action} has no derivatives, so its variation produces no term on $\partial D$. Finally, the face conditions also restrict the gauge transformations, which makes everything gauge invariant as well. 

\paragraph{Sliding the defect to symmetry boundary.} Now let $t_0 \rightarrow 0$. The piece $\mathcal{B}^L_{\text{face}}$ disappears, the whole face carries $\iota^\ast b = 0$, and all the fields are the $(a_R,b_R)$ ones; so, we drop the subscript. The defect action \eqref{eq: defect-action} now lives on $\Sigma_2 \times\{0\}$ and, combines with \eqref{eq: sym-boundary-edge} into \eqref{eq: dual symmetry boundary action}. The corner action of $\mathsf{D}_R$ was written in terms of $b_L = s^{-1}a$. Altogether,
\begin{equation}\label{eq: action after fusing}
\begin{aligned}
    S = \frac{i}{2\pi}\int_{M_3} a \, db - \frac{iR}{2\pi s}\int_{\Sigma_2 \times\{0\}} X\, da &+ \frac{i}{2\pi}\int_{\Sigma_2 \times \{0\}} b\, a 
     + \frac{i}{2\pi} \int_{C_{\text{Top}}} \Psi\, dX \\ 
    & + \frac{iR}{2\pi s} \int_{C_{\text{Top}}} X \,a + \frac{1}{4\pi} \int_{\Sigma_2\times\{1\}} b \,\star b,
\end{aligned}
\end{equation}
where $X$ and $\Psi$ are the scalars we added on the edge and the corner respectively, with the following gauge transformations
\begin{equation}
    \begin{aligned}
        X \rightarrow X - (s/R)\lambda_b, \\
        \Psi \rightarrow \Psi + (R/s)\lambda_a.
    \end{aligned}
\end{equation}
These are the modified gauge transformations obtained from the original $X \rightarrow X - R^{-1}\lambda^L_a$, $\Psi \rightarrow \Psi + R\lambda^L_b$, rewritten with the glued gauge parameters.

Let us vary \eqref{eq: action after fusing}. Instead of writing all the terms, we only write down terms on $\Sigma_2 \times \{ 0\}$,
\begin{equation}
\begin{aligned}
    & \delta \Big( -\frac{iR}{2\pi s}\int_{\Sigma_2\times\{0\}} X da \Big) \\
    &= - \frac{iR}{2\pi s} \int_{\Sigma_2 \times\{0\}} \big( \delta X da - dX \delta a \big) - \frac{iR}{2\pi s} \int_{C_{\text{Top}}} X \delta a,
\end{aligned}    
\end{equation}
combining the contribution coming from $M_3$, the surface terms are same as in \eqref{eq: dual-symmetry boundary-variation}, giving the boundary conditions, 
\begin{equation}\label{eq: vertical boundary conditions}
\boxed{    b\big|_{\Sigma_2\times\{0\}} = -\frac{R}{s} dX, \qquad da \big|_{\Sigma_2\times\{0\}} = 0 , \qquad \oint a \in \frac{2\pi s}{R} \mathbb{Z} }
\end{equation}
These are the same boundary conditions  as in \eqref{eq: dual-symmetry boundary-conditions}. The corner term $-\frac{iR}{2\pi s}\int_{C_{\text{Top}}} X\delta a$ on the other hand, cancels a piece in the variation of the counterterm $\frac{iR}{2\pi s}\int_{C_{\text{Top}}} X a$, whose variation is 
\begin{equation}
    \frac{iR}{2\pi s} \int_{C_{\text{Top}}} (\delta X \,a + X\, \delta a).
\end{equation}
 The full variation at the corner is therefore
\begin{equation}
    \delta S\big|_{C_{\text{Top}}} = \frac{i}{2\pi}\int_{C_{\text{Top}}} \Big[ \delta \Psi dX + \delta X \Big( \frac{R}{s}  a - d\Psi \Big) \Big] ,
\end{equation}
and the corner equations of motion are
\begin{equation}\label{eq: vertical corner eom}
  \boxed{  \Psi: ~ dX\big|_{C_{\text{Top}}} = 0 , \qquad X:~ \frac{R}{s} a\big|_{C_{\text{Top}}} = d\Psi. }
\end{equation}
These are the Dirichlet corner equations \eqref{eq:corner-eom-D} with $Rb_L = (R/s) a$. Note how the compatibility condition now arises. The face fixes $\iota^\ast b = 0$, and the symmetry boundary fixes $b = -(R/s) dX$; at the corner, where the two meet, we must have $dX|_{C_{\text{Top}}} = 0$, which is the first equation in \eqref{eq: vertical corner eom}. This is a sanity check for our calculations. The boundary condition is still Dirichlet for $X$, and it becomes Neumann only once we describe the symmetry boundary in terms of the dual edge mode.

\paragraph{Dualizing the edge mode in the presence of the corner.} We now repeat the dualization \eqref{eq: dual edge mode}-\eqref{eq: dual edge symmetry} on the surface $\Sigma_2\times\{0\}$ with boundary $C_{\text{Top}}$. We use fact that $a$ is closed on the boundary and its periods are quantized to impose, 
\begin{equation}
    a = -\frac{s}{R} dX',
\end{equation}
where $X'$ is the dual edge mode we introduced in \eqref{eq: dual edge mode}. Now if we impose this condition on \eqref{eq: action after fusing}, 
\begin{equation}\label{eq: modified boundary}
    \frac{iR}{2\pi s} \int_{\Sigma_2 \times \{ 0 \} } X\,da + \frac{i}{2\pi} \int_{\Sigma_2 \times \{ 0 \} } b\,a \longrightarrow \frac{i(s/R)}{2\pi} \int_{\Sigma_2 \times \{ 0 \} }b\, dX'  ,
\end{equation}
and the corner terms, 
\begin{equation}
    \frac{i}{2\pi} \int_{C_{\text{Top}}} \Psi \,dX + \frac{iR}{2\pi s} \int_{C_{\text{Top}}}   X\,a \longrightarrow \frac{i}{2\pi} \int_{C_{\text{Top}}} \Psi dX -\frac{i}{2\pi} \int_{C_{\text{Top}}} X dX' .
\end{equation}
Now, we can do field redefinition on the right hand side, 
\begin{equation}
    \Psi_1 : = -\Psi, \qquad \Psi_2 := -X,
\end{equation}
in terms of which the corner terms become, 
\begin{equation}
    \frac{i}{2\pi} \int_{C_{\text{Top}}} \Psi_1 d\Psi_2 + \frac{i}{2\pi} \int_{C_{\text{Top}}} \Psi_2 dX',
\end{equation}
we are still missing one term from the Neumann configuration specified in \eqref{eq: D and N configurations}. That additional term will come by doing an integration by parts in \eqref{eq: modified boundary}. Combining that, we can write down the complete transformation of the corner configurations, 
\begin{equation}
\boxed{  \frac{i}{2\pi} \int \Psi dX + \frac{iR}{2\pi s} \int   Xa  \longrightarrow    \frac{i}{2\pi} \int \Psi_1 d\Psi_2 + \frac{i}{2\pi} \int \Psi_2 dX' + \frac{i(s/R)}{2\pi} \int X' b. }
\end{equation}
Just for a consistency check, one can compute the corner equations of motion. These would be, 
\begin{equation}
    d\Psi_2 =0, \qquad dX' = d\Psi_1 \qquad (s/R) b = d\Psi_2.
\end{equation}
Now, if we specialize to $s=1$,we can say that (see Figure~\ref{fig: vertical defect} for an illustration.), 
\begin{equation}
 \boxed{   \mathcal{T}_1 [D] \otimes \mathsf{D}_R = \mathsf{N}_{1/R}. }
\end{equation}
Combining this with $\mathcal{T}_1 [D] \otimes \mathcal{B}^{(R)}_{\text{Sym}} = \mathcal{B}^{(1/R)}_{\text{Sym}}$, we can reduce the sandwich, and we will obtain the free boson theory at radius $1/R$ with the Neumann boundary condition. This is a boundary SymTFT realization of the fact that T-duality exchanges Dirichlet and Neumann boundary conditions.

\subsection{Horizontal placement: open condensation defect }

As of now, we have placed the defect $\mathcal{T}_s$ vertically in the SymTFT. Now, we are going to place the defect horizontally, so that it will be anchored between the symmetry boundary and the physical boundary. 
\begin{equation}
    H = \gamma \times [0,1] \subset M_3, \qquad \gamma \subset \Sigma_2.
\end{equation}
This configuration was named as open condensation defect in \cite{Argurio:2024ewp}. After the interval compactification, $H$ becomes a line along $\gamma$ in the two dimensional theory. Pushing $H$ onto the topological face amounts to the fusion of this surface with the boundary condition on the face. The horizontal defect divides $M_3$ into two regions, $M_{\text{down}}$, which does not touch the face, and the collar $M_{\text{up}}$, which contains the face and the two corners. We denote the fields by $(a,b)$ in $M_{\text{down}}$ and $(a',b')$ in $M_{\text{up}}$. 

To begin with, we fix our conventions, like we did in \eqref{eq: gluing conditions}. We also set $s = 1$ throughout this part\footnote{For $s\neq \pm 1$ the physical boundary condition would change across $\gamma \times\{1\}$, from $b = i \star a$ to $b = i s^{-2}\star a$, and the end of the defect on the physical boundary would be a non-topological interface between bosons of different radii. We do not consider this. The case $s=-1$ differs from $s=1$ by charge conjugation.},
\begin{equation}\label{eq: horizontal gluing}
 \boxed{   a' = b, \qquad b' = a \qquad \text{on } H, \qquad S_{\mathcal{T}_1} = \frac{i}{2\pi}\int_H a\wedge b. }
\end{equation}
The boundary of $H$ consists of the two curves
\begin{equation}
    \partial H = \underbrace{\big( \gamma \times \{0\} \big)}_{:=J_0} \cup \underbrace{\big( \gamma \times \{1\} \big)}_{:=J_1}.
\end{equation}
They split the symmetry boundary and the physical boundary into two halves which we are going to call $\mathcal{B}^{\text{up/down}}$.

\subsubsection{The ends of the defect} 

\paragraph{The end on the physical boundary ($J_1$).} On the physical boundary, the region below $J_1$ carries the action $\frac{1}{4\pi}\int b\wedge \star b$ and above that the collar carries $\frac{1}{4\pi}\int b'\wedge \star b'$. With $s=1$ and the gluing conditions, the two impose the same boundary condition, $b = i\star a$ on both sides of $\gamma\times\{1\}$, and the two boundary actions agree on-shell. Neither action contains derivatives, and neither does $S_{\mathcal{T}_1}$, so no term is produced on $J_1$ by the variation, and the gauge parameters vanish on the physical boundary, so no term is produced there by a gauge transformation either. The end of $H$ on the physical boundary is therefore transparent, which is consistent with what we saw earlier, the physical boundary is self--dual. The corner $C_{\text{Phys}}$ is also not affected.

\paragraph{The end on the symmetry boundary ($J_0$).} We start from the Dirichlet configuration in the collar, since the collar is where the face and the corners are: the face carries $\iota^\ast a' = 0$, the corner $C_{\text{Top}}$ carries the Dirichlet corner theory \eqref{eq: D and N configurations} written in the primed fields, with the collar edge mode $X'$, 
\begin{equation}
    S_{C_{\text{Top}}} = \frac{i}{2\pi} \int_{C_{\text{Top}}} \Psi dX' + \frac{iR}{2\pi} \int_{C_{\text{Top}}} X'b',
\end{equation}
and both parts of the symmetry boundary carry the boundary condition $\mathcal{B}^{(R)}_{\text{Sym}}$ in their own variables.
\begin{equation}\label{eq: horizontal symmetry boundaries}
\begin{aligned}
     \mathcal{B}^{\text{down}}_{\text{Sym}} &:~ -\frac{iR}{2\pi}\int X db, \quad a = -R dX , \\
     \mathcal{B}^{\text{up}}_{\text{Sym}} &:~ -\frac{iR}{2\pi}\int X' db', \quad a' = -R dX'.
\end{aligned}
\end{equation}
So, unlike the vertical case, boundary conditions of the SymTFT change across $J_0$, and we have to ask what lives on this junction; in other words, a junction action is needed. 

We determine it by the same logic we used for the corners: we compute the terms produced on $J_0$ by the variation of the actions on the two sides, and ask for the minimal set of terms on $J_0$ which makes the variational problem well posed \textit{without} imposing any condition on the bulk fields beyond the boundary conditions themselves.

Varying \eqref{eq: horizontal symmetry boundaries}, produce the terms, 
\begin{equation}\label{eq: horizontal line terms}
    -\frac{iR}{2\pi}\int_{J_0} X \delta b \qquad \text{and} \qquad -\frac{iR}{2\pi}\int_{C_{\text{Top}}} X'\delta b' + \frac{iR}{2\pi}\int_{J_0} X' \delta b' = -\frac{iR}{2\pi}\int_{C_{\text{Top}}} X'\delta b' + \frac{iR}{2\pi}\int_{J_0} X' \delta a,
\end{equation}
where the relative sign of the two $J_0$ terms comes from our choice of orientations, and we used $b' = a$. The variation also produces a  corner contribution on $C_{\text{Top}}$ because $\mathcal{B}^{\text{up}}_{\text{Sym}}$ has two disconnected boundaries, 
\begin{equation}
    \partial \mathcal{B}^{\text{up}}_{\text{Sym}} = C_{\text{Top}} \cup \{-J_0 \}, 
\end{equation}
where we have reversed the orientation as we have chosen a positive orientation for the boundary of $\mathcal{B}^{\text{down}}_{\text{sym}}$.

In the expression above, the two terms on $J_0$ are of the same type, and are traded for terms proportional to $\delta X$ and $\delta X'$ by adding local counterterms on $J_0$,
\begin{equation}
    \frac{iR}{2\pi}\int_{J_0} X b - \frac{iR}{2\pi}\int_{J_0} X' a ,
\end{equation}
i.e. we are just writing both symmetry boundary actions in the ``$dX\wedge b$'' form instead of as in equation \eqref{eq: sym-boundary-edge}. What is left on $J_0$ after this is
\begin{equation}
    \frac{iR}{2\pi} \int_{J_0} \big( \delta X b - \delta X' a \big), 
\end{equation}
and if nothing else lived on $J$ these would impose $a|_{J_0} = b|_{J_0} = 0$: the two boundary conditions would be joined by an interface -- totally reflective interface -- on which no line can pass.

Moving on, consider adding
\begin{equation}\label{eq: identification term}
    S_{J_0}^{\text{id}} = -\frac{i\kappa}{2\pi}\int_{J_0} X' dX, \qquad \delta S^{\text{id}}_{J_0} = -\frac{i\kappa}{2\pi}\int_{J_0} \big( \delta X' dX - dX' \delta X \big) ,
\end{equation}
where $X'$ is another $2\pi$-periodic scalar. So, the total variation on $J_0$ becomes
\begin{equation}\label{eq: junction variation}
    \delta S\big|_{J_0} = \frac{i}{2\pi}\int_{J_0} \Big[ \delta X \big( R b + \kappa dX' \big) - \delta X' \big( R a + \kappa dX \big) \Big] .
\end{equation}
The junction equations are therefore 
\begin{equation}
    b|_{J_0} = - (\kappa/R) dX', \qquad a|_{J_0} = -(\kappa/R) dX.
\end{equation}
On the other hand, the boundary conditions \eqref{eq: horizontal symmetry boundaries} restricted to $J_0$ already say $a|_{J_0} = -R dX$ (from the down side) and $b|_{J_0} = -R dX'$ (from the collar side). The two sets of equations coincide precisely, when
\begin{equation}\label{eq: kappa}
    \kappa = R^2 .
\end{equation}
With this value the complete junction action, including the two counter terms, is
\begin{equation}\label{eq: junction action}
\boxed{  S_{J_0} = \frac{iR}{2\pi} \int_{J_0} X b - \frac{iR}{2\pi} \int_{J_0} X' \big( a + R dX \big).  }
\end{equation}
It can be checked (see Appendix~\ref{app: appendix 1} for detailed derivation) that the junction is gauge invariant. So, \eqref{eq: junction action} is a consistent, gauge invariant junction with no degrees of freedom of its own.
 
The last term of \eqref{eq: junction action} couples two $2\pi$-periodic scalars with the coefficient $R^2$. Whether such a coupling is well defined, and whether the winding sectors of the two edge modes fit together across $J_0$, are global questions, and they are where the radius enters. The local manipulations of the next paragraph do not depend on them, so we first see what the line does to the boundary condition, and return to the global questions afterwards.

\begin{figure}[ht]
    \centering
    \begin{tikzpicture}[scale=1.35]
        \fill[blue!5] (0,0) rectangle (7,2.8);
        \draw[line width=2.2pt, red!65!black] (0,0) -- (0,2.8);
        \draw[line width=2.2pt, blue!65!black] (7,0) -- (7,2.8);
        \draw[line width=2.2pt, orange!55!black] (0,2.8) -- (7,2.8);
        \draw[line width=1.8pt, green!45!black, dashed] (0,1.85) -- (7,1.85);
        \fill[green!45!black] (0,1.85) circle (2.6pt);
        \fill[green!45!black] (7,1.85) circle (2.6pt);
        \fill[red!65!black] (0,2.8) circle (3.2pt);
        \fill[blue!65!black] (7,2.8) circle (3.2pt);
        \node[green!45!black] at (3.5,1.6) {$\mathcal{T}_1$ on $H = \gamma\times[0,1]$};
        \node[green!45!black, anchor=east] at (-0.15,1.85) {$J_0$};
        \node[orange!55!black, anchor=south] at (3.5,2.95) {$\mathcal{B}_{\text{face}}:~\iota^\ast a' = 0$};
        \node[anchor=south east] at (-0.1,3.05) {$C_{\text{Top}}:\{\Psi\}$};
        \node[anchor=south west] at (7.1,3.05) {$C_{\text{Phys}}$};
        \node[anchor=south west] at (7.1,1.85) {$J_1$};
        \node[anchor=east] at (-0.15,2.35) {$a' = -RdX'$};
        \node[anchor=east] at (-0.15,0.9) {$a = -RdX$};
        \node[anchor=west] at (7.15,1.4) {$\mathcal{B}_{\text{Phys}}$};
        \node at (3.5,2.3) {$M_{\text{up}}$:~$(a',b')$};
        \node at (3.5,0.9) {$M_{\text{down}}$:~$(a,b)$};
        \node at (-0.3,0.3) {(a)};
        \begin{scope}[yshift=-4.6cm]
        \fill[blue!5] (0,0) rectangle (7,2.8);
        \draw[line width=2.2pt, red!65!black] (0,0) -- (0,2.8);
        \draw[line width=2.2pt, blue!65!black] (7,0) -- (7,2.8);
        \draw[line width=2.2pt, orange!55!black] (0,2.8) -- (7,2.8);
        \fill[red!65!black] (0,2.8) circle (3.2pt);
        \fill[blue!65!black] (7,2.8) circle (3.2pt);
        \node[orange!55!black, anchor=south east] at (4.9,2.95) {$\mathcal{B}_{\text{face}}:~\iota^\ast b = 0$};
        \node[anchor=south west] at (-1.85,3.05) {$C_{\text{Top}}:\{\Psi_1 = -\Psi,\ \Psi_2 = -X'|_{C_{\text{Top}}}\}$};
        \node[anchor=south west] at (7.1,3.05) {$C_{\text{Phys}}$};
        \node[anchor=east] at (-0.15,1.4) {$\mathcal{B}_{\text{Sym}}$};
        \node[anchor=west] at (7.15,1.4) {$\mathcal{B}_{\text{Phys}}$};
        \node at (3.5,1.1) {$\mathcal{T}_1$ pushed onto the face ($R=1$)};
        \node at (-0.3,0.3) {(b)};
        \end{scope}
    \end{tikzpicture}
    \caption{(a) The horizontal defect $H = \gamma\times[0,1]$, which after the interval compactification is the T-duality line of the boson at radius $R$ along $\gamma$. Its end on the physical boundary is transparent; its end $J$ on the symmetry boundary is the junction \eqref{eq: junction action} between $\mathcal{B}^{(R)}_{\text{Sym}}$, described by $X$, and $\mathcal{T}_1\cdot\mathcal{B}^{(R)}_{\text{Sym}}$, described by $X'$. (b) At the self dual radius we fuse the defect with the face producing Neumann configuration \eqref{eq: horizontal result}.}
    \label{fig: horizontal defect}
\end{figure}
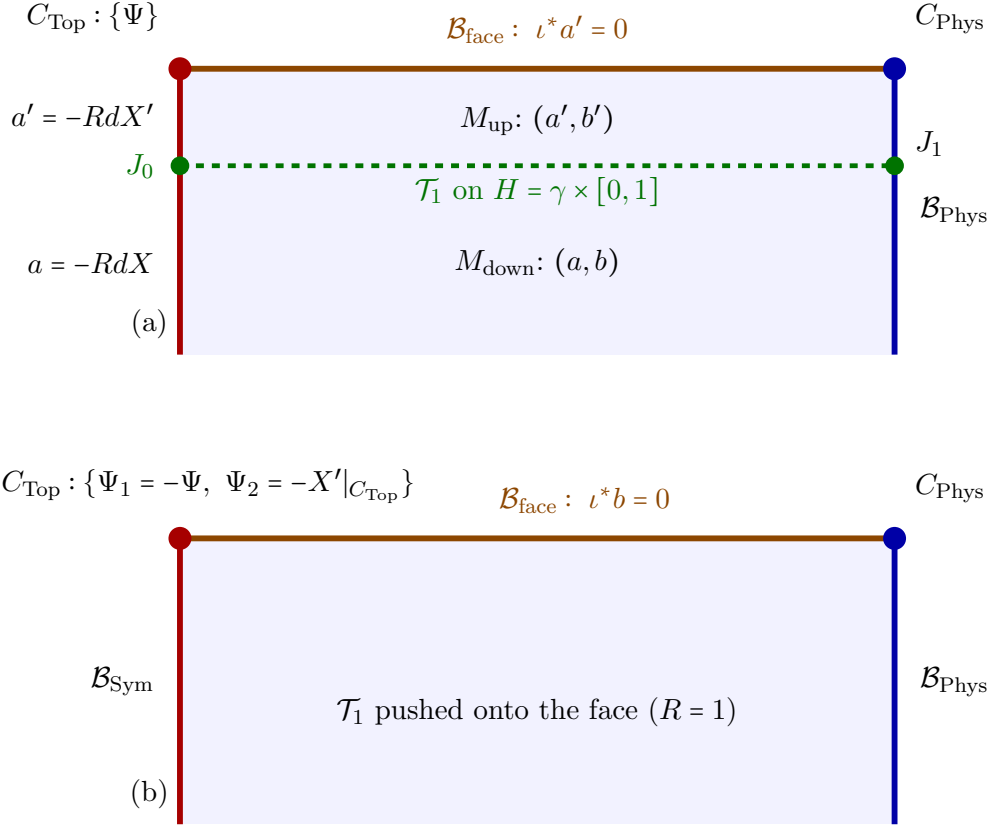

\paragraph{Fusing the defect with $\mathcal{B}_{\text{face}}$.} We now fuse $\mathcal{T}_1$ with the face. The collar shrinks to zero width and the following things happen (see Figure~\ref{fig: horizontal defect}).
\begin{itemize}
    \item The face condition changes. Its condition $\iota^\ast a' = 0$ changes to $\iota^\ast b = 0$. the face has become of Neumann type. 
    \item The defect action becomes a term on the face, $\frac{i}{2\pi}\int_{\mathcal{B}_{\text{face}}} a\wedge b$, which vanishes because $\iota^\ast b = 0$. 
    \item The junction $J_0$ merges with $C_{\text{Top}}$, and the corner counterterm $\frac{iR}{2\pi}\int_{C_{\text{Top}}} X' b'$ cancels against the junction counterterm $-\frac{iR}{2\pi}\int_{J_0} X' a$ of \eqref{eq: junction action}, since $b' = a$.
    \item  $\mathcal{B}^{\text{down}}_{\text{Sym}}$ now extends all the way to $C_{\text{Top}}$, together with the first term of \eqref{eq: junction action}, which is precisely its corner counterterm $\frac{iR}{2\pi}\int_{C_{\text{Top}}} X b$.
    \item The edge mode $X'$ in the collar, survives only through its value on $C_{\text{Top}}$, which becomes a corner scalar.
\end{itemize}
Keeping all of these in mind and collecting the Dirichlet corner term $\frac{i}{2\pi}\int \Psi dX'$, the new configuration after the fusion is
\begin{equation}\label{eq: horizontal final configuration}
    \mathcal{B}^{(R)}_{\text{Sym}}, \qquad \iota^\ast b\big|_{\mathcal{B}_{\text{face}}} = 0, \qquad S_{C_{\text{Top}}} = \frac{i}{2\pi} \int_{C_{\text{Top}}} \Big( \Psi dX' - R^2 X' dX \Big) + \frac{iR}{2\pi}\int_{C_{\text{Top}}} X b,
\end{equation}
hence $X'$ has eventually become a corner scalar, as emphasized earlier. Doing an integration by parts, 
\begin{equation}
    S_{C_{\text{Top}}} = \frac{i}{2\pi}\int_{C_{\text{Top}}} \big( \Psi + R^2 X \big) dX' + \frac{i}{2\pi}\int_{C_{\text{Top}}} X b,
\end{equation}
and setting $R=1$, we get
\begin{equation}\label{eq: horizontal self dual corner}
    S_{C_{\text{Top}}}\big|_{R=1} = \frac{i}{2\pi}\int_{C_{\text{Top}}} \big( \Psi + X \big) dX' + \frac{i}{2\pi}\int_{C_{\text{Top}}} X b,
\end{equation}
which is the Neumann corner action of \eqref{eq: D and N configurations} at $R=1$ with the following field redefinitions
\begin{equation}\label{eq: horizontal result}
\boxed{ \mathcal{T}_1[H] \otimes  \mathcal{B}_{\text{face}} : \qquad \mathsf{D}_{1} \longmapsto \mathsf{N}_{1}, \qquad \Psi_1 := -\Psi , \qquad \Psi_2 := -X'\big|_{C_{\text{Top}}}. }
\end{equation}
After the interval compactification this is the self-dual boson with the Neumann edge mode theory \eqref{eqn:Neumann_TFT} on $\partial\Sigma_2$. Thus, the invertible T-duality line, fused onto the topological face with Dirichlet polarization, gives the Neumann polarization, and by the same computation with the roles of $a$ and $b$ exchanged, it maps the Neumann boundary back to the Dirichlet one. For $R \neq 1$, \eqref{eq: horizontal final configuration} is a two edge mode theory of the general type studied in Section~\ref{sec: free boson boundary}, with a coupling $R^2$ between the two edge modes; what it describes is decided by the global questions we postponed, to which we now turn.

\subsubsection{Global properties of the junction $J_0$.}

The preceding analysis determines the local properties of the junction. However, this is not the complete answer yet, because both $X$ and $X'$ are compact scalars. Their winding sectors, and the corresponding quantization of the bulk holonomies, must also be compatible across $J_0$.

It is already visible in the last term of \eqref{eq: junction action}, 
\begin{equation}\label{eq: horizontal compact mixed term}
    -\frac{iR^2}{2\pi} \int_{J_0} X'dX.
\end{equation}
For $X$ and $X'$ being unrestricted $2\pi$-periodic scalars, the level $R^2$ must be an integer. 

On the other hand, we can imagine restricting the allowed winding numbers of $X$ and $X'$ and see what conditions that imposes on $R^2$.  For example, $X'\rightarrow X'+2\pi$ changes \eqref{eq: horizontal compact mixed term} by
\begin{equation}
\Delta S = -2\pi i R^2 w_X, \qquad w_X:=\frac{1}{2\pi}\oint_{J_0}dX\in\mathbb Z.
\end{equation}
Thus the exponentiated action is unchanged only if $R^2w_X\in\mathbb{Z}$ \footnote{One should not conclude from this that the horizontal defect exists only when $R^2$ is an integer.  At a general radius, the open condensation defect comes with a projection onto precisely those winding sectors for which the coupling is well defined. }. Hence, together with the boundary condition ($a = - R dX$) and the compactness of $X$ these imply
\begin{equation}\label{eq: horizontal down global data}
\oint_{J_0} a = -2\pi Rw_X \in 2\pi R  \mathbb{Z},  \qquad \oint_{J_0} b \in \frac{2\pi}{R} \mathbb{Z}.
\end{equation}
The second statement follows from the sum over the winding sectors of $X$ in $\frac{iR}{2\pi}\int Xdb$. On the collar side using the same argument, $X'$ and the primed fields, gives
\begin{equation}\label{eq: horizontal up global data}
\oint_{J_0}a' = -2 \pi R w_{X'} \in 2\pi R  \mathbb{Z}, \qquad \oint_{J_0}b' \in\frac{2\pi}{R}\mathbb{Z}, \qquad w_{X'}:=\frac{1}{2\pi}\oint_{J_0} dX'\in\mathbb{Z}.
\end{equation}
Using the gluing conditions $a'=b$ and $b'=a$,
\begin{equation}
\oint_{J_0} a\in 2\pi R \mathbb{Z} \cap \frac{2\pi}{R} \mathbb{Z}, \qquad \oint_{J_0} b\in 2\pi R\mathbb{Z} \cap \frac{2\pi}{R}\mathbb{Z}.
\end{equation}
Equivalently, the allowed winding sectors obey
\begin{equation}\label{eq: horizontal winding compatibility}
\boxed{ R^2w_X\in\mathbb Z, \qquad R^2w_{X'}\in\mathbb Z. }
\end{equation}
This is the global content of the junction. At $R=1$, the conditions \eqref{eq: horizontal winding compatibility} impose no restriction: every winding sector of $X$ and $X'$ is allowed. The junction is therefore transparent both locally and globally.  This is why the horizontal defect becomes an invertible line of the theory precisely at the self-dual radius, and why its fusion with the Dirichlet face gives the Neumann corner in \eqref{eq: horizontal self dual corner}. If $R^2=p/q$, with $\gcd(p,q)=1$, then the condition tells us that we need to project $X$ and $X'$ on to sectors whose winding is a multiple of $q$.  This effectively changes the periodicities of the scalars.

\paragraph{Beyond the self-dual radius.} For $R \neq 1$ the conditions \eqref{eq: horizontal winding compatibility} are not empty: the junction only admits the winding sectors of $X$ and $X'$ for which $R^2 w$ is an integer, and projects out all the others. This is how the boundary SymTFT sees the well known fact that the T-duality line of the compact boson is an invertible symmetry only at the self-dual radius, and a non-invertible line otherwise \cite{Thorngren:2021yso, Niro:2022ctq, Argurio:2024ewp}. For the purposes of this paper the self-dual case is all we need, and we will not develop the fusion of the non-invertible line with the face here. Let us only record what the corner theory \eqref{eq: horizontal final configuration} suggests, since it will connect with the next section. The coupling $-\frac{iR^2}{2\pi}\int X' dX$ between the two compact corner scalars has the non-integer level $R^2$, and it is only consistent on the winding sectors singled out by \eqref{eq: horizontal winding compatibility}\footnote{At a rational squared radius $R^2 = p/q$ these are the sectors in which the windings are multiples of $q$. On them $X'|_{C_{\text{Top}}}$ can be written as $q$ times a $2\pi$-periodic scalar $\varphi$, the level becomes the integer $p$, and \eqref{eq: horizontal final configuration} turns into one of the two edge mode theories of the general form \eqref{general lagrangian}, with $(\Psi_1,\Psi_2) = (\varphi,\Psi)$, $v = p$ and $k_1 = q$. According to the analysis of the subsection \ref{operator algebras} such a theory is a non-simple boundary condition of Neumann type, a superposition of Neumann boundaries, which is what one expects from the action of a non-invertible line on a simple boundary. We do not pursue this here.}. 

For irrational $R^2$ the only surviving sector is the one in which neither $X$ nor $X'$ winds around $\partial\Sigma_2$. In that sector the corner scalar $X'$ can be treated as if it were non-compact, and \eqref{eq: horizontal final configuration} takes the form of an edge mode theory with one compact and one non-compact mode, precisely of the type we are going to study in Section \ref{sec: Noncompact edge modes}. As we will see there, non-compact edge modes are what produces boundary conditions smeared over the position of the brane. It is therefore tempting to speculate that at irrational radius the T-duality line maps the Dirichlet boundary not to a single Neumann boundary but to a boundary smeared over the dual circle. This is also what one would guess physically: the position on the dual circle is conjugate to the winding, and a junction which projects out every winding sector erases exactly the information that would fix it. We leave a proper analysis of the T-duality line away from the self-dual radius for the future.

\section{Non-compact modes and continuous spectrum}\label{sec: Noncompact edge modes}

The edge mode theories we have used as of now consist only of $U(1)$-valued scalars ($\Psi_i$) localized on the boundary. We also saw (in subsection \ref{operator algebras}) that the spectrum of $N$ compact edge modes is discrete for any $N$. The next obvious generalization is to use non-compact ($\mathbb{R}$-valued) modes localized on the boundary. 

To motivate the use of non-compact modes, beyond Dirichlet and Neumann boundary conditions, a proposed set of boundary conditions called the Friedan-Janik boundaries \cite{Friedan:99, Janik:2001hb}, and their boundary states are written as
\be\label{Friedan boundary}
||F(x)\rrangle=\mathcal{C}_{x}\sum_{J=0}^{\infty}P_{J}(x)||[J,J]\rrangle,
\ee
where $\mathcal{C}_{x}$ is a normalization factor, $P_{J}(x)$ is the $J$-th Legendre polynomial, $x$ lies in the range $-1\leq x\leq 1$, and $||[J,J]\rrangle$ are the Ishibashi states corresponding to the following primary operators
\be
U_{[J,J]}(z,\bz)=\mathcal{N}_{J}V_{J}(z)\overline{V}_{J}(\bz)\;\;\;\text{with}\;\;\; V_{J}(z)=\lp\int \frac{du}{2\pi}e^{-i\sqrt{2}X(z+u)}\rp^Je^{i\sqrt{2}JX(z)},
\ee
where $\mathcal{N}_{J}$ is a normalization factor. The states in \eqref{Friedan boundary} have a continuous boundary spectrum \cite{Friedan:99, Janik:2001hb, Cai:2025ukh}. Although these boundaries have pathologies\footnote{They have an infinite boundary entropy \cite{Tseng:2002ax} and violate the cluster condition if written in the RCFT language \cite{Cai:2025ukh}}, in this section, our goal would be to realize these boundary states using the edge mode technology. However, as we saw in subsection \ref{operator algebras}, the compact modes give us only a discrete boundary spectrum, and thus, in order to recover the Friedan-Janik states we will now also include non-compact ($\mathbb{R}$-valued) edge modes $(\Phi_a)$ in our boundary action, as suggested in \cite{Arbalestrier:2025jsg}.

\subsection{General edge action} The most general edge mode action that we will consider is
\begin{align}
    S^{\text{non-cpt}}_{\text{edge mode}} = \frac{i}{2\pi} \int_{\partial \mathcal{M}} \Big[ (v_{i}\Psi_{i} + u_a \Phi_a) dX + \frac{1}{2} k_{ij}\Psi_{i}\;d\Psi_{j} + \frac{1}{2} \wt{k}_{ab} \Phi_a&\; d\Phi_b + \wh{k}_{ia} \Psi_i ~ d\Phi_a\nonumber\\
    \label{non compact action}
    &+\;\al_0 dX+  \al_{i}\;d\Psi_{i} \Big],
\end{align}
where $k$ and $\wt{k}$ are antisymmetric matrices, and while $k$ is integral, $\wt{k}$ need not be. Similarly, $v_i$ is an integral vector while $u_i$ is real. Let's say we have $N$ compact and $M$ non-compact modes.

With this normalization, the boundary equations of motion are
\begin{align}
        \Psi_i &: v_i dX + k_{ij} d\Psi_j + \wh{k}_{ia}d\Phi_a =0 , \\
        X &: R^2 \star dX -i v_i d\Psi_i -i u_a d \Phi_a =0, \\
        \Phi_a &: u_a dX + \wt{k}_{ab} d \Phi_b - \wh{k}_{ia} d\Psi_i =0. 
\end{align}
We can write these equations in a simpler form by introducing the following objects
\be
dY=R^2\star dX,\efill V=\begin{pmatrix}
    v\\u
\end{pmatrix},\efill q=\begin{pmatrix}
    \Psi\\\Phi
\end{pmatrix},\efill \Om=\begin{pmatrix}
    k&\wh{k}\\
    -\wh{k}^T&\wt{k}
\end{pmatrix},
\ee
to get the following equations of motion
\begin{align}
\label{eq:non compact modes eoms neat form 1}
dY=iV^T dq,\\
\label{eq:non compact modes eoms neat form 2}
V dX+\Om dq=0.
\end{align}
We can also write the action \eqref{non compact action} in a more compact form as
\be\label{non compact action neat form}
S^{\text{non-cpt}}_{\text{edge mode}}=\frac{i}{2\pi}\int_{\p\mathcal{M}}\lp V^Tq dX+\hlf q^T\Om dq+\al_0 dX+\al_i\Psi_i\rp.
\ee
Now, the quantization constraints are
\begin{align}
    \Psi_{i}&:v_iX + k_{ij}\Psi_j + \wh{k}_{ia}\Phi_a - \al_i \in 2\pi\mathbb{Z},\\
    X&: R^2\wt{X}-v_i\Psi_i-u_a\Phi_a-\al_0\in 2\pi\mathbb{Z},
\end{align}
with $\Phi_{a}$ having no quantization constraint as it is a non-compact edge mode.

Now, we can analyze the equations (\ref{eq:non compact modes eoms neat form 1}) and (\ref{eq:non compact modes eoms neat form 2}) in two different cases i.e. $\Om$ being invertible or not (i.e.~singular). In this discussion we'll only consider the local equations of motion, so we don't actually need to worry about which components are integral or not, just whether $\Om$ is invertible as a real $(N+M)\times (N+M)$ matrix.  If $\Om$ is invertible (which can only happen if the number of edge modes is even), then~(\ref{eq:non compact modes eoms neat form 2}) implies
$$
dq=-\Om^{-1}VdX,
$$
and putting this result in (\ref{eq:non compact modes eoms neat form 1}) implies
$$
dY=-iV^T\Om^{-1}VdX=0,
$$
where the last equality follows due to the fact that $\Om$ is antisymmetric. Since $dY=R^2\star dX$, we have a Neumann boundary. Since $\Om$ is an invertible $(N+M)\times (N+M)$ matrix, any $ N+M$ dimensional vector is in its image set and hence, $V\in \text{Im}\;\Om$. This point will become important in a moment.

Now, let's consider the case where $\Om$ isn't invertible. If $V$ is in the image of $\Om$, say $V=\Om p$ for some vector $p$.  In that case~\eqref{eq:non compact modes eoms neat form 2} tells us that $dq=p\,dX+dr$, where $dr$ is in the kernel of $\Om$, and plugging into~\eqref{eq:non compact modes eoms neat form 1} we get
\begin{equation}
    dY=ip^T\Om\left(p\,dX+dr\right)=0.
\end{equation}
Hence we again get a Neumann boundary condition.  On the other hand, if $V$ is not in the image of $\Om$, then we can find a vector $z$ which is in the kernel of $\Om$ and satisfies $z^TV\ne 0$.  In that case, contracting~\eqref{eq:non compact modes eoms neat form 2} with $z^T$ implies that $dX=0$, so we get Dirichlet.  

\subsection{Specific cases}

Let's study some cases in more detail. If $N=0,M=1$ i.e. no compact modes and one non-compact mode, we have
\begin{equation}\label{single non compact, action}
    S^{\text{non-cpt}}_{\text{edge mode}} = \frac{1}{2\pi} \int_{\partial \mathcal{M}} \Big[ u \Phi dX +\;\al_0 dX\Big].
\end{equation}
Since $\Om$ vanishes, $V$ is not contained in the image of $\Om$, which implies that we would get a Dirichlet boundary (it can be confirmed by using the equations of motion). The quantization constraint for $X$ implies
\be\label{single non compact, winding constraint}
\Phi(x)=\frac{R^2}{u}\wt{X}(x)-\frac{\al_0}{u}\text{ mod }\frac{2\pi\Z}{u},
\ee
but since there are no quantization constraints coming from the compact edge modes, the value of $X$ isn't fixed to a given value. We can make sense of these boundaries by calculating the OPE of the holomorphic stress tensor with vertex operator $:e^{ip\Phi(x)}:$ with $p\in\R$ using \eqref{single non compact, winding constraint} to get
\be
T(z):e^{ip\Phi(x)}:\;\sim \frac{ p^2R^2}{u^2}:e^{ip\Phi(x)}:+\cdots.
\ee
As $p/u\in\R$, the form of the conformal weights appearing here is
$$
h=R^2s^2\text{ with }\;\;s\in\R.
$$
Since the position of $X$ isn't specified by the equations of motion (and $dX=0$), this can be a hint that we have a smeared Dirichlet boundary here. To investigate this possibility, we consider the following smeared Dirichlet state
\be\label{smeared Dirichlet state}
||\mathfrak{D}\rrangle=\int_0^{2\pi}dx_0 ||D(x_0)\rrangle,
\ee
and calculate the cylinder amplitude of this state with itself to get
\begin{align}
\llangle \mathfrak{D}|q^H|\mathfrak{D}\rrangle&=\int_0^{2\pi}dx_0\int_0^{2\pi}dx'_0\llangle D(x_0)|q^H|D(x'_0)\rrangle\nonumber\\
\label{smeared Dirichlet overlap}
&=\int_0^{2\pi}dx_0\int_0^{2\pi}dx'_0\sum_{m\in\Z}\chi^{U(1)}_{R^2\lp m-\frac{x_0-x'_0}{2\pi}\rp^2}.
\end{align}
Since $x_0$ and $x'_0$ both vary from $0$ to $2\pi$, the set of conformal weights varies through all non-negative real numbers. In addition, in Appendix~\ref{app: Path integral}, we derive the density of states that arises in the annulus partition function of the boundaries described by the edge mode action with a single non-compact edge mode (and no compact edge modes), and we get the same spectrum as \eqref{smeared Dirichlet overlap}. This gives strong reason to believe that \eqref{single non compact, action} produces a smeared Dirichlet boundary condition. We can also calculate the density of states corresponding to different conformal weights in \eqref{smeared Dirichlet overlap} by expressing the state \eqref{smeared Dirichlet state} in terms of a particular Friedan-Janik boundary and use the results in \cite{Cai:2025ukh}. We get the following result
\be
||\mathfrak{D}\rrangle=\frac{2\pi}{\mathcal{C}_{1}\sqrt{R\sqrt{2}}}||F(1)\rrangle
\ee
$$
\Rightarrow \llangle \mathfrak{D}|q^H|\mathfrak{D}\rrangle=\frac{4\pi^2}{R|\mathcal{C}_{1}|^2\sqrt{2}}\llangle F(1)|q^H|F(1)\rrangle=\frac{4\pi^2}{R}\int_0^{\infty}\frac{dh}{\sqrt{h}}\chi_{h}(\wt{q})
$$
\be
\Rightarrow \rho(h)=\frac{4\pi^2}{R\sqrt{h}}.
\ee
Therefore, we have a continuous density of states, as expected.

Similarly, we can consider the $N=M=1$ case, i.e., one compact and one non-compact mode. In this case, $\Om$ is invertible, and we are guaranteed to get a Neumann boundary condition, which can be checked using the equations of motion. Once again, the fixed value of $\wt{X}$ isn't provided by the quantization constraints. The equations of motion also provide us with the following relations
\be
\Psi(x)=\frac{uX(x)}{\wh{k}}+\text{constant},\efill \Phi(x)=-\frac{vX(x)}{\wh{k}}+\text{constant},
\ee
and using these relations, we can calculate the OPE of the vertex operator $e^{in\Psi+ip\Phi}$ with $n\in\Z$ and $p\in\R$ to get
\be
T(z):e^{in\Psi(x)+ip\Phi(x)}:\sim\frac{1}{R^2}\lp\frac{nu-pv}{\wh{k}}\rp^2\frac{:e^{in\Psi(x)+im\Phi(x)}:}{(z-w)^2}+\cdots,
\ee
which again gives us all possible non-negative real conformal weights which can be generated by varying $p$. The boundary in question behaves like a smeared Neumann boundary (i.e. smeared on the dual circle where $\wt{X}$ lives) which is given as
\be
||\mathfrak{N}\rrangle=\int^{\frac{2\pi}{R^2}}_{0}d\wt{x}_0||N(\wt{x}_0)\rrangle.
\ee
This boundary is proportional to the Friedan-Janik boundary $||F(-1)\rrangle$ as
\be\label{smeared Neumann state}
||\mathfrak{N}\rrangle=\frac{2\pi}{R^2\;\mathcal{C}_{-1}}\sqrt{\frac{R}{\sqrt{2}}}||F(-1)\rrangle,
\ee
and its overlap with itself generates a continuum of non-negative conformal weights
\be\label{smeared neumann overlap}
\llangle \mathfrak{N}|q^H|\mathfrak{N}\rrangle=\int^{\frac{2\pi}{R^2}}_{0}d\wt{x}_0\int^{\frac{2\pi}{R^2}}_{0}d\wt{x}'_0\sum_{n\in\Z}\chi_{\frac{1}{R^2}\lp n+\frac{R^2}{2\pi}(\wt{x}_0-\wt{x}'_0)\rp^2}(\wt{q}).
\ee
These results strongly suggest that the boundary condition described by the $N=1, M=1$ case includes the smeared Neumann boundary. The density of states for \eqref{smeared neumann overlap} however, is divergent. It can be seen using the general density of states expression in \cite{Cai:2025ukh} with $\t_1=\t_2=\pi$ which gives
\be\label{smeared neumann density of states}
\rho(h)\propto \frac{1}{R^3\sqrt{h}}K(1)=\infty,
\ee
where $K(u)$ is the complete elliptic integral of the first kind, defined as
\be
K(u)=\int_0^1\frac{dt}{\sqrt{(1-t^2)(1-u^2t^2)}},
\ee
and $K(1)$ is infinite.

So, we see that using non-compact edge modes we can get only the Dirichlet or Neumann boundaries (depending on whether $V$ is in the kernel or image of $\Om$) and these conditions can give rise to smeared Dirichlet and Neumann boundaries, which are like the Friedan-Janik boundaries $||F(1)\rrangle$ and $||F(-1)\rrangle$ respectively, but the states $||F(x)\rrangle$ for $-1<x<1$ still elude the edge mode action \eqref{non compact action}.

\section{Multiple bulk fields}\label{sec: multiple bulk}

In this section, we would like to use our edge mode techniques to explore the boundary states of a CFT with multiple bulk free bosons. Such CFTs are called torus CFTs and/or multi-component free boson CFTs, and their boundary CFT is studied in the context of string theory and condensed matter physics \cite{Oshikawa:2010kv, condensed1, condensed2, condensed3, condensed4}. Although this topic may feel like a digression from the main thread of the paper, it provides a natural extension of several of the ideas developed in the preceding sections. Our aim is not to give a comprehensive treatment, but rather to record a few preliminary observations that we hope to investigate more systematically in future work.

The action for multiple compact bosons (say $D$ bosons) living on a torus $T^D$ is
\be
S^{\text{Torus}}=\frac{1}{4\pi}\int_{\mathcal{M}}\lp G_{AB}\;dX^A\wedge \star dX^B+iB_{AB}\;dX^A\wedge dX^B\rp=-\frac{i}{4\pi}\int_{\mathcal{M}} dX^A\wedge d\wt{X}_A
\ee
where $G_{AB}$ and $B_{AB}$ are the (constant) metric an anti-symmetric $B$ field respectively, and
\be
-id\wt{X}_A=G_{AB}\star dX^B+i B_{AB}dX^B
\ee
The variation of the action with respect to $X^A$ gives the following boundary variation
\be
\d S^{\text{Torus}}_{\text{boundary}}=-\frac{i}{2\pi}\int_{\p\mathcal{M}}\d X^A\wedge d\wt{X}_A
\ee
The edge mode action, with $N$ compact edge modes, that we consider is
\be
S^{\text{Torus}}_{\text{edge modes}}=\frac{i}{2\pi}\int_{\p\mathcal{M}}\lp V_{iA}\Psi^idX^A+\hlf k_{ij}\Psi^id\Psi^j+\al_A dX^A+\al_i d\Psi^i\rp
\ee
The boundary equations of motion for $\Psi^i$ and $X^A$ are
\be
V_{iA}dX^A+k_{ij}d\Psi^j=0\efill d\wt{X}_A=-V_{iA}d\Psi^i
\ee
which can be written in the matrix notation as
\begin{align}\label{torus CFT eom 1}
VdX+kd\Psi=0\\
\label{torus CFT eom 2}
d\wt{X}=-V^Td\Psi
\end{align}
and we can repeat the arguments similar to the previous section to conclude that if $k$ is invertible, then
\be
d\wt{X}-\mathcal{F}dX=0\text{ where }\mathcal{F}=V^Tk^{-1}V
\ee
Note that the combination $V^Tk^{-1}V$ doesn't vanish this time, as we have more than one bulk field now. These are fluxed Neumann boundaries. Moreover, if $k$ is singular, then let the columns of a matrix $Z$ be the basis vectors of the kernel of $k$, and multiply \eqref{torus CFT eom 1} by $Z^T$ to get 
\be\label{torus CFT, Dirichlet boundaries}
Z^TV dX=0\Rightarrow Z^TV X=c\;\;(\text{mod } 2\pi\Z^N)
\ee
where $c$ is a constant vector. This equation sets $\text{dim }(\text{ker} (k))$ (which we will refer to as $d$ from now) linear combinations of $X^A$ to constants, but not all of them may be independent constraints, and solution space of these conditions is a subset of $T^D$ which can be seen as a Dirichlet brane. Let the basis vectors of the tangent space of this brane be the columns of a matrix $T$. This would imply that
\be\label{torus CFT, tangent constraint}
Z^TVT=0
\ee
To investigate the structure of the solutions of \eqref{torus CFT, Dirichlet boundaries} let's define $X=BY$ where $B\in GL(D,\Z)$ and then \eqref{torus CFT, Dirichlet boundaries} implies
\be\label{Dirichlet boundary after transformation}
\lp AZ^TVB\rp Y=c'
\ee
where $A\in GL(N,\Z)$ and $c'=Ac$. The periodicity of all $Y^A$s is still $2\pi$ as $B^{-1}$ is an integer matrix. Let's choose $A$ and $B$ such that $AZ^TVB$ is equal to the Smith normal form of $Z^TV$. This form has vanishing off-diagonal entries, and the diagonal entries (which are integers) can be zero or nonzero. If the rank of $Z^TV$ is $r$, then we would have $r$ non-zero diagonal entries and $\text{min}(d,N)-r$ zero diagonal entries. Therefore, \eqref{torus CFT, Dirichlet boundaries} impose $r$ independent conditions, and thus, the Dirichlet boundary conditions fix the values of $r$ coordinates, and the dimension of the Dirichlet brane(s) is $\text{min}(d,N)-r$.

Now, let the non-zero diagonal entries of the Smith normal form be $\la_{A}$ for $A$ in some set $\mathfrak{Z}$. The conditions \eqref{Dirichlet boundary after transformation} then become
$$
\la_{A}Y^{A}=c'\;\;(A \text{ isn't summed over})
$$
We see that for any $\la_{A}>1$, this condition would produce multiple copies of the Dirichlet brane, but for all $\la_A=1$, there is only one connected Dirichlet brane. In fact, the number of connected components of the Dirichlet brane is simply
\be
\text{Number of connected components}=\prod_{A\in\mathfrak{Z}}\la_A
\ee
If all $\la_A=1$, then the lattice generated by the rows of $Z^TV$ is called primitive, and if any $\la_A>1$, then we get a non-primitive lattice. Therefore, a connected Dirichlet brane is obtained only if the row lattice of $Z^TV$ is primitive. To see a simple example, recall the example with a single bulk field and a single compact edge mode. In this case, $V=(v)$, and $k=0$ which implies $Z=(1)$. This leads to the following equation of motion
\be
vdX=0
\ee
as can be seen from \eqref{compact edge modes eom psi1}. We argued earlier that $v>1$ corresponds to multiple D0 branes on the target space circle. This happens because the Smith normal form of $Z^TV$ is just $(v)$ and it is primitive only if $v=1$.

Let's come back to the analysis of the equations of motion \eqref{torus CFT eom 1} and \eqref{torus CFT eom 2}. We can write $d\Psi$ as a sum of a vector in $\text{ker}(k)$ and a vector in $\text{Im} (k)^\vee$, the dual of the image of $k$, as
\be\label{torus CFT, dpsi breakdown}
d\Psi=d\Psi_{\text{Im}^\vee}+d\Psi_{\text{ker}}=-k^{-1}_{\text{Im}}VdX+Zd\zeta
\ee
where $d\zeta$ is a vector of parameters for $\text{ker}(k)$, and $d\Psi_{\text{Im}^\vee}$ is determined using \eqref{torus CFT eom 1}. The inverse $k^{-1}_{\text{Im}}$ is defined only on the image space of $k$ \footnote{This inverse can be defined because $k$ won't map two different vectors to the same vector if we restrict our vectors to be strictly in $\text{Im}(k)$.  This inverse is not necessarily defined over the integers, only over the rationals}. Using \eqref{torus CFT, dpsi breakdown} in \eqref{torus CFT eom 2}, we get 
\be
d\wt{X}-V^Tk^{-1}_{\text{Im}}VdX=-V^TZd\chi
\ee
If we multiply this equation from the left by $T^T$, the right-hand side vanishes because of \eqref{torus CFT, tangent constraint}, and we get
\be\label{torus CFT, Neumann boundaries}
T^{T}\lp d\wt{X}-\mathcal{F}dX\rp=0
\ee
where this time $\mathcal{F}$ is defined similarly but with $k^{-1}_{\text{Im}}$ instead of $k^{-1}$. Therefore, for a singular $k$, the boundary conditions are given by \eqref{torus CFT, Dirichlet boundaries} and \eqref{torus CFT, Neumann boundaries}, which describe Dirichlet and (fluxed) Neumann boundaries, respectively. If $k$ is invertible, its kernel is empty and hence, there are no Dirichlet boundaries conditions, and we only have the fluxed Neumann boundary conditions.

\section{Discussion and Outlook}

In this paper, we have used the edge mode technology for the compact free boson CFT to derive the family of Dirichlet and Neumann boundary conditions. We also show that using the most general action with compact edge modes that we consider, we are able to get only Dirichlet or Neumann states, depending on how many edge modes we have. 

We then placed the same system inside the boundary SymTFT of the compact boson. The Dirichlet and Neumann edge mode theories were recovered as the corner theories that make the topological face compatible with the symmetry boundary, the two polarizations of the face corresponding to the two boundary conditions. This description separates the data of the boundary condition from the data of the physical theory; the radius sits only on the symmetry boundary, while the face and the corners know only which polarization has been chosen. We then used the T-duality defect of the SymTFT to relate the two. Placed parallel to the symmetry boundary, the defect maps the Dirichlet configuration at radius $R$ to the Neumann configuration at radius $1/R$, and it does so by turning the boundary value of the original edge mode into the second Neumann corner mode; placed across the sandwich, it becomes the T-duality line of the theory, which at the self-dual radius is an invertible symmetry that maps the Dirichlet edge mode theory to the Neumann one when fused with the boundary.

Although we were able to get smeared Dirichlet and Neumann boundaries using the non-compact edge modes, which correspond to the Friedan-Janik states $||F(1)\rrangle$ and $||F(-1)\rrangle$ respectively, the generic Friedan-Janik states $||F(x)\rrangle$ for $-1<x<1$ still elude this edge modes technology. It would be an interesting direction to explore whether these boundaries can be obtained through this edge mode technology, and since we still don't know what the meaning of these boundaries is in terms of the field $X(z,\bz)$, the edge mode technology may provide a way to fill this gap.

Another future direction, relevant to the BCFT of the compact free boson, that can be pursued here is to obtain the Gaberdiel-Recknagel boundaries \cite{GaberdielRecknagel:2001} that exist only at rational radii. To achieve these boundaries, we need to bring in the other two $SU(2)$ currents, i.e., $J^{1}$ and $J^2$, into the picture, because the gluing condition for these boundaries includes these currents. One potential way to obtain these boundaries at the self-dual radius is to include marginal deformations generated by these currents in the boundary action, along the lines of \cite{Hasselfield:2005nv}, which would give us the aforementioned currents in the equations of motion. If the Gaberdiel-Recknagel boundaries can be obtained from the edge mode technology, then it can also provide a route to obtain Friedan-Janik boundaries, because they can be obtained from the Gaberdiel-Recknagel boundaries in the irrational radius limit as shown in \cite{GaberdielRecknagel:2001}.

Lastly, it would be interesting to push this technology to other CFTs. Since the techniques that we are using are heavily dependent on the existence of a Lagrangian, the obvious candidates for future directions would be the Ising CFT, the SCFT with a single boson and fermion, the Liouville CFT, and the WZW models. Each of these CFTs comes with challenges of its own (e.g., dealing with fermions in the Ising CFT, and non-linear terms in the Liouville theory), and it would be a fruitful exercise to push this edge mode technology to the aforementioned CFTs.

\subsection*{Acknowledgments}

S.R. thanks Xingyang Yu and Pinak Banerjee for valuable discussions and for collaboration on a related project. The authors would like to gratefully acknowledge support from the Simons Center for Geometry and Physics, Stony Brook University, at which some of the research for this paper was performed. S.R. thanks the organizers of Lotus \& Swamplandia 2026 and Global Categorical Symmetries 2026 for the hospitality during part of this work. S.R. thanks IHP Paris, the Department of Physics at Virginia Tech and the Theory group at UPenn for the hospitality during part of this work. 

\appendix

\section{Density of states from the edge mode action}\label{app: Path integral}
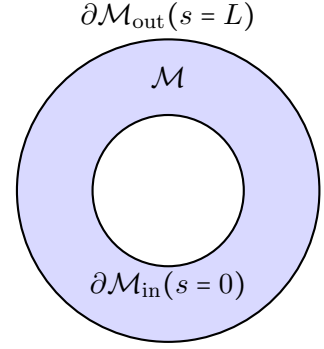
\begin{wrapfigure}{r}{0.3\textwidth}
    \centering
    \begin{tikzpicture}
\fill[blue!15] (0,0) circle (2cm);

\fill[white] (0,0) circle (1cm);

\draw[thick] (0,0) circle (2cm);
\draw[thick] (0,0) circle (1cm);

\node at (0,-1.25) {$\partial \mathcal{M}_{\mathrm{in}} (s=0)$};
\node at (0,1.5) {$\mathcal{M}$};
\node at (0,2.3) {$\partial \mathcal{M}_{\mathrm{out}}(s=L)$};
    \end{tikzpicture}
    \caption{The annulus for the path integral computation}
    \label{fig:annulus}
\end{wrapfigure}
We will calculate the partition function with the bulk action and a single non-compact edge mode action on an annulus with the bulk labeled as $\mathcal{M}$ and the boundaries labeled as $\p\mathcal{M}_{\text{in}}$  and $\p\mathcal{M}_{\text{out}}$ as shown in Figure~\ref{fig:annulus}. We do this computation by calculating the following path integral
\be
\mathcal{Z}=\int \mathcal{D}X\;\mathcal{D}\psi_{\text{in}}\mathcal{D}\psi_{\text{out}}e^{-S},
\ee
where the overall action is given as
\begin{align}
S&= S_{\text{bulk}}+S_{\text{bd.}}\nonumber\\
&=\frac{R^2}{4\pi}\int_{\mathcal{M}} dX\wedge \star dX+\frac{i}{2\pi}\int_{\p\mathcal{M}_{\text{out}}} \Phi_{\text{out}}\;dX\nonumber\\
&-\frac{i}{2\pi}\int_{\p\mathcal{M}_{\text{in}}} \Phi_{\text{in}}\;dX.
\end{align}
The inner boundary action gets a minus sign because of the orientation reversal. The coordinate on this annulus can be taken as $w=e^{s+i\t}$ where $0\leq s\leq L$. In the $(s,\t)$ coordinates, the mode expansion of $X(s,\t)$ and $\Phi_{a}(\t)$ are (where $a=$ in or out)
\be\label{X mode expansion}
X(s,\t)=x_{0}(s)+w\t+\sum_{n\neq 0}x_{n}(s)e^{in\t},
\ee
\be
\Phi_{a}(\t)=\Phi_{a,0}+\sum_{n\neq 0}\Phi_{a,n}e^{in\t},
\ee
where $x_{-n}(s)=x^{*}_{n}(s)$. Using these mode expansions, the boundary action becomes
\be\label{boundary action simplified}
S_{\text{bd.}}=i\lp\Phi_{\text{out},0}-\Phi_{\text{in},0}\rp w-\sum_{n\neq 0}n\lp \Phi_{\text{out},-n}x_{n}(L)-\Phi_{\text{in},-n}x_{n}(0)\rp.
\ee
If we integrate over $\Phi_{\text{out},-n}$ and $\Phi_{\text{in},-n}$ for $n\neq 0$, we get factors proportional to
$$
\d\lp nx_{n}(L)\rp\d\lp nx_{n}(0)\rp
$$
which sets the oscillator parts of $X$ equal to zero on the boundaries. Integrating over $\Phi_{\text{out},0}$ and $\Phi_{\text{in},0}$ sets $w=0$. Therefore, $X$ is a constant on the boundaries, which means
$$
X(0,\t)=x_{\text{in}}\efill X(L,\t)=x_{\text{out}}
$$
However, these boundary values aren't fixed by the edge mode action. This suggests that we have smeared Dirichlet conditions on the two boundaries, as we will shortly confirm. We will now perform the bulk path integral. In our convention, we have
\be
\star dX=-\p_{\t}X\;ds+\p_{s}Xd\t
\ee
which leads to the following expression for the bulk action
\be
S_{\text{bulk}}=\frac{R^2}{4\pi}\int_{0}^{L}ds\;\int_{0}^{2\pi}d\t\lp (\p_{s}X)^2+(\p_{\t}X)^2\rp
\ee
Using the values of $X$ at the boundaries, the zero mode $x_{0}(s)$ can be written as
\be
x_{0}(s)=x_{\text{in}}+\frac{x_{\text{out}}-x_{\text{in}}+2\pi N}{L}s+y(s)\;\;\;\;N\in\Z
\ee
where $y(s)$ is a function such that $y(0)=y(L)=0$. Using this result, with the mode expansion \eqref{X mode expansion}, the bulk action becomes
\be\label{bulk action simple form}
S_{\text{bulk}}=\frac{2\pi^{2}R^2}{L}\lp N+\frac{x_{\text{out}}-x_{\text{in}}}{2\pi}\rp^2+\frac{R^2}{2}\int_{0}^{L}ds(\p_{s}y)^2+R^2\int_{0}^{L}ds\;\sum_{n=1}^{\infty}\lp |\p_{s}x_{n}|^2+n^2|x_{n}|^2\rp
\ee
Let's consider the $x_{n}$ part of the action in \eqref{bulk action simple form}. For each individual $n$, this action is just like the harmonic oscillator action in Euclidean signature. If we split $x_{n}$ into its real and imaginary parts as $x_{n}=a_{n}+ib_{n}$, then the result of the integration over $a_{n}$, with the boundary conditions $a_{n}(L)=a_{n}(0)=0$, is (e.g., see \cite{mackenzie2000pathintegralmethodsapplications})
\be\label{an result}
\int_{a_{n}(0)=0}^{a_{n}(L)=0}\mathcal{D}a_{n} e^{-R^2\int_{0}^{L}ds\;(\dot{a}^2_{n}+n^2a^2_{n})}=\sqrt{\frac{R^2n}{\pi \sinh (nL)}}
\ee
and a similar result holds for $b_{n}$ integration, implying that the result for $x_{n}$ integration is the square of the result in \eqref{an result}. Now, we can do the following manipulation
\be
\prod_{n=1}^{\infty}\frac{R^2n}{\pi \sinh (nL)}=\prod_{n=1}^{\infty}\frac{2R^2n}{\pi}\frac{e^{L/12}}{1-e^{-2nL}}
\ee
where we performed the Riemann zeta function regularization $1+2+...=-1/12$. Defining $q=e^{-2L}$, we see that
\be
\prod_{n=1}^{\infty}\frac{R^2n}{\pi \sinh (nL)}=\frac{1}{\eta(q)}\prod_{n=1}^{\infty}\frac{2R^2n}{\pi}
\ee
where $\eta(q)$ is the Dedekind $\eta$ function.

For the zero-mode part of \eqref{bulk action simple form}, we need to integrate over $y(s),\;x_{\text{in}}$, and $x_{\text{out}}$, and sum over $N$. The $y(s)$ integration can be done easily by taking the $n\rr 0$ limit of the result in \eqref{an result}. Moreover, since $q=e^{-2L},$ we have $\wt{q}=e^{-2\pi^2/L}$ and thus, the zero mode part of the path integral becomes
\be
\sqrt{\frac{R^2}{2\pi L}}\int_{0}^{2\pi} dx_{\text{out}}\int_{0}^{2\pi} dx_{\text{in}}\sum_{N\in\Z}\wt{q}^{R^2\lp N+\frac{x_{\text{out}}-x_{\text{in}}}{2\pi}\rp^2}
\ee
Using the fact that $\eta(q)$ transforms as
$$
\frac{1}{\eta(q)}=\sqrt{\frac{L}{\pi}}\frac{1}{\eta(\wt{q})},
$$
we get the following expression for the complete path integral (up to an $ L$-independent normalization constant)
\be\label{density of states from Dirichlet}
\mathcal{Z}\propto\int_{0}^{2\pi} dx_{\text{out}}\int_{0}^{2\pi} dx_{\text{in}}\sum_{N\in\Z}\frac{\wt{q}^{R^2\lp N+\frac{x_{\text{out}}-x_{\text{in}}}{2\pi}\rp^2}}{\eta(\wt{q})}
\ee
 which is exactly the expression for the overlap of two smeared Dirichlet states.

If we replace $\Phi$ with a compact edge mode $\Psi$, write our edge mode integrand as $-X d\Psi$ using integration by parts, and have an additional term in the edge mode action, which is $\al\;d\Psi$ (just like in \eqref{Dirichlet_Family_TFT}) the mode expansion of $\Psi_{a}$ will get a winding term as well, which is $m_{a}\t$, and we will get the following extra terms in the boundary action
\be
-m_{\text{out}}\lp x_{\text{out}}+2\pi w-\al_{\text{out}}\rp+m_{\text{in}}\lp x_{\text{in}}+2\pi w-\al_{\text{in}}\rp.
\ee
We see that summing over the edge winding modes will set $x_{a}=\al_{a}$ mod $2\pi$ where $a=\{\text{in, out}\}$, which agrees with \eqref{eq: Dirichlet family generation}, and the integrals over $x_{\text{out}}$ and $x_{\text{in}}$ in \eqref{density of states from Dirichlet} will go away, giving us Dirichlet states on the two boundaries, instead of smeared Dirichlet states.

\section{Gauge invariance of the junction}\label{app: appendix 1}

In this appendix, we check for the gauge invariance of on the junction. We have contributions coming from the bulk, defect, symmetry boundary pieces, and junction action itself. Recall that $(\lambda_a,\lambda_b)$ and $(\lambda'_a,\lambda'_b)$ denote the gauge parameters in $M_{\mathrm{down}}$ and $M_{\mathrm{up}}$, respectively,
\begin{equation}\label{eq: horizontal bulk gauge transformations}
\begin{aligned}
    a&\rightarrow  a + d\lambda_a,
    & \qquad b &\rightarrow b+d\lambda_b,
    \\
    a' &\rightarrow a'+ d\lambda'_a,
    & b'&\rightarrow b'+d\lambda'_b.
\end{aligned}
\end{equation}
Preservation of the gluing condition, $a'=b$ and $b'=a$ on $H$ requires
\begin{equation}\label{eq: horizontal gauge gluing}
    \lambda'_a|_H=\lambda_b|_H, \qquad
    \lambda'_b|_H=\lambda_a|_H,
\end{equation}
The transformations of the edge modes on the symmetry boundary $(X, X')$, 
\begin{equation}\label{eq: horizontal edge gauge transformations}
    \delta X=-\frac{1}{R}\lambda_a, \qquad
    \delta X'=-\frac{1}{R}\lambda'_a  =-\frac{1}{R}\lambda_b
\end{equation}
We set $c:=i/(2\pi)$ and treat $\lambda_a$ and $\lambda_b$ separately. The $\lambda_a$ transformation.
\begin{equation}
    \delta a=d\lambda_a, \qquad \delta b'=d\lambda_a, \qquad \delta X=-\frac{\lambda_a}{R},
\end{equation}
while $b,a'$ and $X'$ are unchanged. First, we vary the junction action itself:
\begin{align}\label{eq: horizontal junction lambda a variation}
    \delta_{\lambda_a}S_{J_0}
    &=cR\int_{J_0}\delta X b
      -cR \int_{J_0} X'\big(\delta a+R d\delta X\big)
      \notag\\
    & = - c\int_{J_0} \lambda_a b.
\end{align}
Next, the defect action changes by
\begin{align}\label{eq: horizontal defect lambda a variation}
    \delta_{\lambda_a}S_{\mathcal T_1}  & = c\int_H d\lambda_a\wedge b  \notag\\
    & = -c\int_H\lambda_adb + c\int_{J_0}\lambda_a b.
\end{align}
We can already see that we do not need to add any extra term on $J_1$; the gauge variation of the defect action vanishes because the allowed gauge parameter vanishes on the physical boundary. 

The $\mathcal{B}^{\text{down}}_{\text{sym}}$ action is not invariant by itself. Its variation gives
\begin{equation} \label{eq: horizontal down boundary gauge variation}
    \delta_{\lambda_a}S^{\text{down}}_{\text{Sym}} =-cR\int_{\mathcal B^{\text{down}}_{\text{Sym}}} \delta X db = c\int_{\mathcal B^{\text{down}}_{\text{Sym}}} \lambda_a db
\end{equation}
whereas the $\mathcal{B}^{\text{up}}_{\text{sym}}$ action is invariant under $\lambda_a$, as $db'$ is unchanged and $X'$ is inert. Finally, the BF action (of the bulk) in the down region gives
\begin{equation}\label{eq: horizontal down bulk gauge variation}
\begin{split}
    \delta_{\lambda_a}S^{\text{down}}_{\text{bulk}}
    &= c\int_{M_{\text{down}}} d\lambda_a  db\\
    &= c\int_H \lambda_a db
      -c\int_{\mathcal B^{\text{down}}_{\text{Sym}}}
          \lambda_a db.
\end{split}
\end{equation}
The relative minus sign in the second line follows because the displayed orientation of the symmetry boundary is opposite to its induced orientation as a component of $\partial M_3$. We can see that terms in equations \eqref{eq: horizontal junction lambda a variation}--\eqref{eq: horizontal down bulk gauge variation} now cancel pairwise. 

In the same spirit, we proceed to check the $\lambda_b$ transformations.
\begin{equation}
    \delta b=d\lambda_b, \qquad \delta a'=d\lambda_b, \qquad \delta X'=-\frac{\lambda_b}{R}
\end{equation}
while $a,b',$ and $X$ are unchanged. The junction variation is
\begin{equation}\label{eq: horizontal junction lambda b variation}
    \delta_{\lambda_b}S_{J_0} = cR\int_{J_0} Xd\lambda_b + c\int_{J_0} \lambda_b \big( a + R dX \big) =c\int_{J_0}\lambda_b a,
\end{equation}
The defect action instead gives
\begin{align}\label{eq: horizontal defect lambda b variation}
    \delta_{\lambda_b}S_{\mathcal T_1} &=c\int_H a\wedge d\lambda_b \notag \\ &=c\int_H\lambda_b da - c \int_{\partial H}\lambda_b a \notag \\  & = c\int_H\lambda_bda -c\int_{J_0} \lambda_b a.
\end{align}
On the upper symmetry boundary, the transformation of $X'$ yields
\begin{equation} \label{eq: horizontal up boundary gauge variation}
    \delta_{\lambda_b}S^{\mathrm{up}}_{\mathrm{Sym}}
    = c\int_{\mathcal B^{\mathrm{up}}_{\mathrm{Sym}}}
       \lambda_bdb',
\end{equation}
while the down region of symmetry boundary is invariant. The BF action in bulk contributes 
\begin{equation}\label{eq: horizontal up bulk gauge variation}
\begin{split}
    \delta_{\lambda_b}S^{\mathrm{up}}_{\mathrm{bulk}} & = c\int_{M_{\mathrm{up}}}d\lambda_b\wedge db' \\ 
    & = -c\int_H\lambda_bda -c\int_{\mathcal B^{\mathrm{up}}_{\mathrm{Sym}}} \lambda_b db'.
\end{split}
\end{equation}
The cancellations are again pairwise, the $H$ terms in \eqref{eq: horizontal defect lambda b variation} and \eqref{eq: horizontal up bulk gauge variation} cancel, the two-dimensional collar terms cancel between \eqref{eq: horizontal up boundary gauge variation} and \eqref{eq: horizontal up bulk gauge variation}, and the defect endpoint is canceled by \eqref{eq: horizontal junction lambda b variation}.

Thus the complete bulk--boundary and the defect--junction system is gauge invariant, and the junction requires no additional gauge degree of freedom.

\printbibliography
\end{document}